\documentclass[preprint]{aastex}
\usepackage{aastexug}   % User guide style customizations 
 
\shorttitle{Richer et al.}
\shortauthors{Main Sequence of M4}

\begin{document}

\title{HST Observations of the Main Sequence of M4$^1$}

\author{H.B. Richer, G.G. Fahlman$^2$, J. Brewer, S. Davis, J. Kalirai}
\affil{Dept. of Physics and Astronomy, University of British Columbia}
\affil{richer/jbrewer/jkalirai@astro.ubc.ca}
\affil{greg.fahlman@nrc-cnrc.gc.ca}

\author{P.B. Stetson}
\affil{National Research Council, Herzberg Institute of Astrophysics
         }
\affil{peter.stetson@nrc-cnrc.gc.ca}

\author{B.M.S. Hansen, R.M. Rich}
\affil{Department of Physics and Astronomy,
           University of California at Los Angeles
           }
\affil{hansen/rmr@astro.ucla.edu}

\author{R.A. Ibata}
\affil{Observatoire de Strasbourg}
\affil{ibata@newb6.u-strasbg.fr}

\author{B. K. Gibson}
\affil{Centre for Astrophysics and Supercomputing, Swinburne University}
\affil{bgibson@astro.swin.edu.au}
 
\author{M. Shara}
\affil{American Museum of Natural History
           }
\affil{mshara@amnh.org}

\altaffiltext{1}
{Based on observations with the NASA/ESA Hubble Space Telescope, obtained at
the Space Telescope Science Institute, which is operated by AURA under NASA
contract NAS 5-26555. These observations are associated with proposals GO-5461 and GO-8679.}
\altaffiltext{2}
{Herzberg Institute of Astrophysics}

\begin{abstract}

 This paper presents new results from a photometric study 
 of the main-sequence stars
 in M4 (NGC\,6121 = GC 1620$-$264), the closest globular cluster
 to the sun.   Multi-field, multi-epoch observations 
 at approximately 1, 2, and 6 core-radii were
 obtained with the WFPC2 camera on the HST through either $F606W$ or $F555W$\,($V$) and 
 $F814W$\,($I$) filters. The multi-epoch observations allowed us to clean the data on the basis of proper motion
and thus separate cluster from field stars or extragalactic objects.
  In all the fields the cluster main sequence can be traced to 
at least $V$ = 27.0 but there remains a trail of stars to the limit of the
data near $V = 30$ in the deepest outer field. There is no evidence that we 
have reached the end of the
hydrogen burning main sequence in any of our fields, however, there is some indication
that very few stars remain to be detected in the deepest data.
 
A study of the scatter about the cluster main sequence yields a surprisingly
small and variable binary fraction; $f_b \simeq 2\%$ 
 in the inner parts of the cluster falling to the $1\%$ range outside. However, 
with one possible exception, no
stars in the 6 core radius field exhibit photometric
variability on timescales of a
few hours through a few days. For the currently visible main sequence stars,
the cluster mass function is very flat ($\alpha =0.1$) in the outer field and 
flattens further in the inner fields suggesting well developed
mass segregation.
The observed variation in the mass function
is broadly
consistent with
isotropic, multi-mass Michie-King
models. Because we have a large sample of white dwarfs in the outer field
we are able to show that the cluster IMF above $0.8M_{\odot}$ was considerably
steeper than the present day mass function for low mass stars.

Two appendicies are included in this contribution. The first one is a detailed
discussion of the techniques used to reduce the data while the second provides
a direct comparison between the cluster
 stars and those belonging to the inner spheroid of the Galaxy. This yields a
 relative distance  between the cluster, $d_c$, and the Galactic
 center, $R_o$, of
 $R_o/d_c = 4.36 \pm 0.13$.
With our subdwarf-based estimate of $d_c=1.73 \pm 0.09$ kpc to M4, we find 
 $R_o=7.5 \pm 0.6$ kpc.

\end{abstract}


\section{Introduction}
        
  The high angular resolution achievable with the Hubble Space 
Telescope offers at least four significant advantages over ground based 
observations of
globular star clusters: (1) decreased blending makes accurate 
photometry to faint levels feasible in crowded cluster cores, 
(2) the improved contrast of the sharper stellar profile against a 
lower background allows photometry of very faint stars,
(3) the sharp stellar profile makes possible the measurement of proper
motions on short time baselines which facilitates separation of cluster
and field objects, and (4) image morphology can be more effectively used to exclude
extragalactic objects.
By exploiting all of these improvements, we have been able to study
two stellar populations, the cluster white dwarfs and the low-mass 
main-sequence stars, which have proven inaccessible 
from the ground (Richer  et al. 1995, 1997, 2002; Hansen  et al. 2002). 
These stars play key roles in the dynamical evolution of the clusters
and provide important data pertaining to the theory of 
stellar evolution and star 
formation in the early universe. In this paper, we will present
the results of our study of
the main sequence stars of Messier 4 (NGC 6121 = C1620-264) derived from multi-field
and multi-epoch 
HST observations under our cycles 4 and 9 proposals (GO-5461, GO-8679) as
well as second epoch archival observations obtained from GO-8153.  
 
The properties of M4 relevant to this study are summarized in Table 1.
A critical review of earlier work on the reddening, distance
modulus and metallicity of M4 is given by Dixon \& Longmore 1993. 
Here, we
adopt a distance modulus of $(m-M)_{V} = 12.51 \pm 0.09$ and a reddening
a $E(V-I) = 0.51 \pm 0.02$, which were derived 
by fitting the subdwarfs listed in Manduschev et al. 1996
to the cluster main sequence as discussed in
Richer  et al. 1997. \footnote{These quantites are for magnitudes defined in the
Johnson-Cousins system. In the current paper all magnitudes are in the natural HST system so that these quantites are slightly different. See Table 1 for
the extinctions used in this photometric system.}

For the preferred value of the ratio $R_V = A_V/E(B-V) = 3.8$ (Richer  et al. 1997 and references therein), these
estimates imply a physical distance to M4 of $1.73$ kpc, which is
in excellent agreement with the astrometric value
of $1.72 \pm 0.14\ $ kpc \cite{Pet95}  
 and with the earlier distance based on 
a Baade-Wesselink analysis of four cluster RR Lyrae variables \cite{LJ90}.
M4 is the globular cluster nearest
to the Sun.
There is evidence for variable reddening across
the entire M4 field \cite{CR90}, \cite{Ly95}. Some
differential reddening was seen in a field which overlaps part of our HST
data (Fahlman et al. (1996b)), but there 
are no obvious effects which may be attributed to
 differential reddening in our photometric data. 
 
The iron abundance quoted in Table 1 is from the spectroscopic
analysis of four giants by Drake  et al. (1994).
The metallicity value preferred by 
Dixon \& Longmore (1993), [Fe/H]$=-1.1 \pm 0.25$, is in good agreement with this value,
as is the mean of all the previous work cited by Drake  et al. (1994), 
[Fe/H]$ = -1.17 \pm 0.31$. Carney (1996), who lists [Fe/H]$ = -1.18$ for
this cluster has reviewed the existing data on the $\alpha$-element
 enhancement in M4, and derived a value of $[\alpha/Fe]=0.30$. 
Subsequent to Carney's review, Carretta \& Gratton (1997) found [Fe/H]$=-1.19 \pm 0.03$ for M4 from three giants. Also Routledge  et al. (1997)
used new Ca~triplet measurements for 17 M4 giants as part of a reanalysis
of the globular cluster metallicity scale, and report [Fe/H]$=-1.27 \pm 0.04$
on the Zinn-West scale, and [Fe/H]$=-1.05 \pm 0.03$ on the Carretta-Gratton
scale. Clearly, the specification of cluster metallicities 
accurate to the precision of individual studies continues to be a challenge.
The space velocity of M4 has been measured recently by Bedin  et al. (2003)
and Kalirai  et al. (2003) and the cluster appears to have a 
chaotic orbit confined closely to the disk, but avoiding the inner 
$\sim 1$ kpc of the Galaxy \cite{Daup96}. Consequently, the
dynamical evolution of the stellar population in 
M4 has probably been strongly affected by tidal interactions with 
the Galactic disk. 

The dynamical parameters listed in Table 1 are based on the 
multi-mass Michie-King (MMK) models described in \S 7 of this paper. The derived scale  
radius differs only slightly from that of Trager  et al. (1993), 
$r_s = 55 \arcsec$. The concentration parameter, $c = log(r_t/r_c) = 1.75$,
where $r_t$ is the (model) tidal radius and $r_c$ refers to the
conventional core radius of the luminous stars in M4. This is 
larger than the Trager  et al. (1993) value, $c=1.55$, but this is entirely
due to the extended tidal radius of our MMK model.  The
half-mass relaxation time, $t_{rh}$, was calculated using the Spitzer
formula as in Trager  et al. (1993) but with a mean stellar mass and 
half-mass radius, $r_h = 6.67\,r_s$, taken from our 
multi-mass King model. The central relaxation time, $t_{rc}$, was 
similarly obtained.  To calculate these numbers, we need to estimate the
cluster mass, and the mass to light ratios at the cluster center, 
$({\cal M}/{\cal L})_0$, as well as the global value, 
${\cal M}/{\cal L})$. For this purpose, we have used the the observed
central velocity dispersion, $v_0\,=\,4.2\ {\rm  km\,s^{-1}}$ 
taken from Pryor and Meylan (1993).
The values listed in Table 1 differ from those in the
Pryor and Meylan (1993) compilation only because we have applied our distance 
modulus and extinction values to the same photometry that they used, and, of
course, are using our MMK model for the relevant calculated quantities.

The small
values of $t_{rh}$ and $t_{rc}$ compared to the 
nominal cluster age of $T =12.7$\,Gyr (see Hansen  et al. 2002) strongly suggests that the stellar population in M4
is well relaxed. Indeed the ratio $T/t_{rh}$ is so high that M4 is expected to
 have experienced a core collapse (Cohn 1980, Murphy \& Cohn 1988) for
 which there is no evidence. Therefore, like the structurally similar 
 cluster M71, studied in detail by Drukier  et al. (1992), M4 is expected to have
 an internal
 energy source capable of preventing core collapse. Primordial binary stars
 are a good candidate for this energy source (C\^{o}t\'{e} \& Fischer 1996),
 and this provides strong
 motivation for examing the cluster binary frequency.
  
M4 has been the subject of a number of previous photometric studies. 
Richer \& Fahlman (1984) obtained the first
color-magnitude diagram with a CCD detector and notable subsequent 
studies are those of Alcaino  et al. (1988), Kanatas  et al. (1995), Richer  et al. (1995, 1997), Alcaino  et al. (1997), Pulone  et al.
(1999), Bedin  et al. (2001), Richer  et al. (2002,) and Hansen  et al. (2002). 
 
The scope of this paper is as follows. Following a detailed account of the data
reduction in \S 2 and the first appendix, we describe the general properties of the cluster 
color-magnitude diagram in \S 3.   
In \S 4, we review the observational
evidence for a photometric binary sequence. The cluster main sequence luminosity
and mass functions are derived in \S 5. In \S 6 we address the question whether we have reached the end of the hydrogen-burning main sequence in M4. In \S 7
we 
interpret the observed mass-segregation  
with static MMK
models and this, with input from the current
state of dynamical modelling for globular star clusters in general, allows us to provide an estimate the cluster IMF. The final
section provides an overview of the paper and discusses the major results. A
second appendix  
discusses the 
spheroid main sequence seen projected behind M4 and here we estimate the ratio of the 
distance to M4 to the distance to the Galactic center. Since the M4 distance is well known this allows us to derive an accurate distance to the Galactic center.
     
\section{Observations, Data Reduction \& Reduced Photometry}

All observations were obtained with the WFPC2 instrument on HST. Three fields
were observed: (a) a field within one core radius  
of the cluster center, (2) a field at about 2~$r_c$ and 
(3) an outer field at about
 6~$r_c$. A schematic map of the cluster showing the position of the three
fields and its division into 4 annuli is given in Richer  et al. (1997 see also Ibata  et al. (1999)).  Table 2 contains the geometry for each annulus. In GO-5461 data were obtained through the
 $F555W$ ($V$) and $F814W$ ($I$) filters in all fields. In GO-8679 very deep
images were
secured in both $F606W$ and $F814W$ in the 6 core radius field only and these
served as the second epoch for the outer field. In the current paper the $F606W$ images are
employed for proper motions only and are not used for any detailed analysis; for these
we used the earlier $F555W$ images for uniformity with the other fields. We retrieved
images in all 3 fields from the HST archives for GO-8153. These were 
restricted to $F814W$ exposures and were the second epoch images for the inner
2 fields.
A summary of the exposures used in this paper is given in Table 3.
The inner two fields were  also observed through the
$F336W$ ($U$) filter in order to separate better the white dwarfs from other
stars. Those data were presented in Richer  et al. (1995, 1997) and are rediscussed only briefly here as they have now been cleaned on the basis of proper motion.

With the data from GO-8679, we quickly discovered that the background was variable with time (see
Figure 1) decreasing as the two months of data collection progressed.
On the frames taken early on in the data collection we found that the background sky was almost a factor of 2 higher than expected. This was due to zodiacal light
as M4 has an ecliptic latitude of $\rm \sim 6^{\circ}$ and in late January the angle between the Sun and M4 only amounted to slightly more than 
$\rm 60^{\circ}$. As time progressed during the 67 days over which these outer field data were secured the Sun angle increased and the situation 
improved. The net effect of the higher background
was to reduce our limiting magnitude by about 0.2 mags.

Our aim in reducing the M4 data sets was to produce the best and deepest photometry possible with the existing data set and to
ascertain whether an individual star was a cluster member based on how it moved
over a given temporal baseline. We achieved this by using cluster stars
to register the frames, and then combined the frames from the 2
different epochs separately. As the frames were registered using the
cluster stars, we expect that cluster stars (aside from internal
motions of $\sim 0.5$ mas/year which is smaller than our measurement error) will appear to remain stationary, whereas non-cluster stars and extragalactic objects
will appear to move on account of the differential motion between the
cluster and the background (which consists mainly of stars in the inner halo of the
Galaxy). This is graphically illustrated in Figure 2.
 
The raw data frames had the standard HST pipeline processing applied to them
and a few additional corrections were also applied, as discussed in Richer et 
al. (1997, 2002). 
The photometry was carried out with the DAOPHOT/ALLSTAR package 
developed by Stetson (1994) with a quadratically varying point-spread 
function. As well the ALLFRAME (Stetson 1994) photometry program, standard IRAF routines, and some
utility programs written in C and C-shell scripts for book-keeping purposes were also employed.

To the best
of our knowledge, this is the first time that DAOPHOT has been used to
make proper motion measurements on WFPC data. Further, since the
processing was slightly non-standard, we have outlined our reduction procedures in some
detail in the first appendix.

In the following Tables (4 - 7) we present our photometry from the M4 fields
arranged by annuli. 
The calibration of the data follows the methodology described by Holzman  et al. (1995). 
By applying the synthetic transformations detailed in Table 10 of that work,
we placed the data on the natural system of HST, that is into $F555W$ and $F814W$ magnitudes. The column headings are $RA$, $DEC$ (both J2000), $F555W$, $F814W$, $\mu_{\alpha}cos(\delta)$, $\mu_{\delta}$, where the latter 2 quantities are the proper motion displacements in RA and DEC in mas/yr measured with respect to an
extragalactic reference frame (Kalirai  et al. 2003).    
  
\section{The Color-Magnitude Diagram}

 The color-magnitude diagrams (CMDs) for stars selected to possess total proper motions appropriate to the cluster (within 5 mas/yr of that of the
mean cluster motion) are shown 
in the ($F555W$, $F555W - F814W$) plane (hereafter referred to as $V, V-I$) in the left hand panels of Figures 3a-d. 
In these diagrams the stars
are plotted in the 4 annuli defined in Richer  et al. 1997.  The mean cluster fiducial over the range $21<V<27$ for stars in the inner two annuli is listed in Table 8. At the upper end of this
range saturation effects are just avoided (except in the outer field where they set in near $F555W = 22$) and at the faint end the main sequence
fiducial ceases to be well defined because of a paucity of stars.
The dispersion in the magnitude differences for stars of known magnitude inserted into the frames in both $I$ and $V$ are shown 
in Fig. 4 for stars in annulus 2. This is a reasonable representation of the
errors for the rather short exposures used in the inner fields.

The most prominent feature 
along the main-sequence, the abrupt drop in star numbers at $V \simeq 27.0$,
is {\it not} an artifact. 
A similar, but less dramatic, feature is apparent in the 
CMD of the metal-poor globular cluster NGC 6397 (King  et al. 1998).
This plunging luminosity function is likely accentuated by mass segregation (see
\S 7) and occurs where the theoretical models (eg those of Montalban  et al. 2000) exhibit 
a change in the slope of the mass-luminosity relation. 
At brighter magnitudes, the CMD appears to broaden again but 
this effect arises predominently from
saturated stellar images ($V \leq 20.5$), 
which lead to large errors in the photometry. 
The individual CMDs show very clearly that the photometric 
completeness
limit is a strong function of radius. The 6 $r_c$ field is entirely 
contained
in annulus 4 and is clearly the deepest data partly by design 
(the exposure 
time through the $F555W$ and $F814W$ filters was considerably longer than in the other fields) and partly 
because scattered light and bright star diffraction spikes are much less of a
limiting factor compared to the inner fields.

\section{Photometric Binaries}

   One unusual characteristic of M4 is that the central two-body
 relaxation time (see Table 1) is much smaller than the apparent age
 of the cluster and therefore the cluster core 
 is expected to be in a state of core
 collapse. However, M4 does not exhibit a collapsed core 
 and, therefore,  
 as noted by [C\^{o}t\'{e} \& Fischer (1996), 
 it may well have a centrally concentrated binary population because 
 such stars are the most plausible source for 
 the energy needed to prevent core collapse. 
 Additional discussion
 of the M4 binary fraction in our data 
 is given by Pryor et al. (1996), Richer et al. (1996a) and Ferdman et al. (2003). 

 A casual inspection of the CMDs of Fig.\,3 suggests the possible presence of
 a few similar-mass photometric binaries located $\approx\!0.75$ magnitudes
 above the cluster
 main sequence. A better look at this can be obtained by inspecting
the CMDs with a fiducial line included shifted by
$\Delta V = -0.75$. Such a diagram is shown in Fig. 5 for each of the separate fields. Comparing the number
 of stars near this line for $V > 20$ (to avoid saturation effects)
with the total number of main sequence stars suggests that no more than
 about $2\%$ of the measured stars are approximately equal-mass binaries 
 (see also Richer  et al. 1996a). The actual measured percentages are 2.2, 1.1, 1.1 and 1.8\% from inner to outer annuli (see Table 9). These must be considered upper limits as
optical binaries are not included in this estimate. Almost identical
percentages were estimated from an examination of the $U, U-I$ color magnitude diagrams in the inner fields. These colors are actually somewhat superior
given their larger color baseline. This very low binary frequency is consistent with the apparent lack of variable stars in the outer
annulus of our data (Ferdman  et al. 2003).

As a consistency check on these conclusions we display in Figure 6 the
results of a simulation of binaries in our fields. The models of Montalban 
 et al., the error distributions shown in Figure 4 and mass functions
with slopes $\alpha = 0$ and $1$ (see \S5) were used to generate the CMDs shown here. The binaries
were chosen at random from the mass function, random errors applied consistent
with those in Figure 4 and, for comparison, a fiducial
sequence 0.75 mags brighter than the main sequence is also included. It is clear from this diagram that binary frequencies in the range 1 - 2\% are
about right for our M4 fields whereas a frequency as high as 5\% is incompatible
with the observed CMDs.

 The distribution of the residual in magnitudes of all cluster stars from the fiducial main sequence for each 
 annulus is given in Fig.\,7. 
 Here we take the entire deviation of a star from the main sequence to be
caused by a deviation in $V$ and plot the histogram of this  
 residual for stars in the magnitude range $V = 20 - 26$. In this diagram negative residuals are main sequence binaries (the star is brighter than the
main sequence fiducial for its color) while the only explanation
for positive residuals (other than non-cluster members) is a white dwarf - main sequence binary.
Inspection of Fig.\,7 suggests that a small 
approximately equal-mass main sequence binary fraction (of order 2\%) may be present in annuli 1 and 2, a smaller fraction (near 1\%) in annulus 3 and annulus
4 exhibits no obvious photometric binaries although there is a non-Gaussian
tail in the residual distribution on the main sequence binary side. Annulus
2 may exhibit some evidence for a small component of white dwarf - main sequence
binaries. 

 The derived binary fractions for the inner fields ($\simeq 2\%$)
 is higher than that indicated elsewhere. This supports (albeit weakly)
the hypothesis of an
 enhanced central binary fraction, but some caution is necessary since the 
 photometry in the 
 central field is also the most affected by the presence of
 diffraction spikes from bright stars and related crowded-field artifacts
 which degrade the photometry.  
 These statistical results need to be verified 
 with radial velocity measurements on the possible binaries.  

 C\^{o}t\'{e} \& Fischer (1996) estimated a 
 binary fraction of $f_b\ \simeq\ 15\%$, after correcting 
 for their detection efficiency, from two detections 
 of radial velocity variations among 33 stars. 
 Their stars were located in the radial distance range:   
 $3{\arcmin}.2 \leq r \leq 16{\arcmin}$. The inner part of their field is
 roughly the same radial distance as our outer field where we derive 
 an estimate of $f_b\ \leq\ 2\%$. This difference is somewhat 
 puzzling, but, in view
 of the uncertainties in both studies, 
 can not be usefully pursued without a great deal more observational effort.
     
\section{The Luminosity and Mass Functions}

  The cluster luminosity function (LF) is obtained by simply counting 
the main sequence stars. This procedure is much improved using proper
motion selection as accounting for the field star contamination is no longer the serious problem that it used to be. The luminosity function itself contains
rather limited information and cannot be compared directly from one
globular cluster to another
because of metallicity effects. The quantity than can be compared is the cluster mass function (MF). 

In this work, we have adopted the policy of making the least possible manipulation to
the observational data themselves and make all the
needed corrections in the theoretical plane. This has the advantage that the
errors in the observational quantities remain as purely statistical without
introducing the problem of error propagation. For example, there are
corrections for completeness effects in the star counting that must be
applied and we have 
incorporated these with the theory. 

Our method of proceeding is thus as follows. From the proper motion selected
cluster main sequence we derive the luminosity function in each annulus.
We then used the theoretical models of Montalban  et al. (2000) and power-law
mass functions to generate theoretical luminosity functions. These functions
were then convolved with the detection efficiency in the data in order that
they could be directly compared with the observations and normalized to contain
the same number of stars as the observations. By this procedure then, there
is no freedom to adjust the functions by arbitrary normalization. By a 
$\chi^2$
process, the best fitting model luminosity function was chosen and the mass
function slope for that luminosity function was then taken as the appropriate
mass function in that annulus.
 
The completeness of our data was determined with 
 very extensive artificial star tests which were described briefly in 
 Richer  et al. 2002. The added stars were drawn from an M4 fiducial sequence
 and 4 trials of 1250 stars each were added into all the chips in all the
fields. The stars were added onto a grid which insured that no artificial stars
 overlapped
with each other. The position of the added star at the intersection
points of this grid was given a small random offset
to insure that stars were not all
added at the same positions in the trials, but the shift
was small compared with the grid and PSF size. This technique (first used
by
Piotto \& Zoccali 1999) allows large numbers of stars to be added into
a single frame while insuring that they do not interfere with each other (ie do not change the crowding statistics in the images).
This guarantees that the corrections thus derived provide accurate
incompleteness statistics. The incompleteness
corrections applied were not a single number for
a given magnitude, but we kept track of all the stars recovered irrespective
at which magnitude they were recovered. From this we derived a 
``correction matrix" with input magnitude along one axis and recovered magnitude
along the other. The diagonal elements of this matrix are the recovery fractions for stars recovered
within the same magnitude bin as input while the off-diagonal ones are the
recovery fractions for stars
recovered either brighter or fainter than input. This matrix was then
multiplied by the theoretical luminosity function for comparison
with the observations. In this way the data remain untouched and all the
corrections were done to the theoretical functions.

Another point to be made about deriving the MF is that the theoretical
models used -- those of Montalban  et al. (2000) or any other models for that matter -- for the theoretical mass-luminosity relation do
not fit the lower main sequences of intermediate metallicity globular clusters very well. This is well illustrated in Bedin  et al. (2001). At this point there
is little that can be done about this other than to compare the MF slopes 
derived with the models available from several groups and to calculate the luminosity functions
in both $V$ and $I$ independently and look for agreement in the 
derived MFs. We discuss these points more below.

Figures 8 and 9 illustrate the observed LFs in the 4 annuli in the
$V$ and $I$ bands together with the best fit derived theoretical
LF which sets the MF in that annulus.  The errors in the observations indicated in Figures 8 and 9 are simply $\sqrt N$ with $N$ being 
the number of stars counted in 0.5 magnitude non-overlapping bins. The MF is characterized by its
power law slope $\alpha$ such that $n(m) \propto m^{-\alpha}$ where $n$ is the number
of stars per unit mass and $m$ is the mass.  In general, the reduced $\chi^2$ values for the fit between the model and the data are rather poor for most of the $V$ LFs, and much better for those done in $I$. This may be a reflection of the difficulty
in producing theoretical spectra of intermediate metallicity cool stars in the $V$-band and the greater success in the redder bands (Bedin  et al. 2001, Delfosse  et al. 2000).
What is, however,  enormously 
reassuring is that the power law slopes derived from the $V$ and $I$ data independently are in such good agreement. For the reasons stated above we
adopt the $I$-band power law MF slopes as the value in each annulus and
list these in Table 10.

To check on the effects of using different theoretical models we explored calculating the MFs using the Padova models (Girardi  et al. 2000) and those of Alexander  et al. (1997). These are less useful for our purposes as they are not
tabulated in the HST bandpasses and the Padova models only extend down to $0.15 M_{\odot}$ and not to the 
the hydrogen burning limit as do the other models. Since all the models were listed in Johnson $V$, in order to carry out the comparison
we transformed $F555W$ to Johnson $V$ (see Richer et al. 1997; Ibata et al. 1999) and compared the data with the tabulated
$\rm mass-M_V$ relation for each of the three sets of models. We are mainly interested here in the {\it difference} between
the derived slopes. For the outer field data (the deepest), restricting the low mass end to be $0.15 M_{\odot}$ so that all 3 sets could be compared, we find that the other models produce slopes that are 0.5 (Padova) and 0.2 (Alexander) steeper than the Montalban models. If, however, we just compare the 2 sets of models that
extend all the way to the hydrogen burning limit so that we can use the full range of the data, we find that the Alexander models produce a slope that has $\alpha$ only 0.1 larger than that of the Montalban models. We conclude from this
that model choices are not a large 
source of systematic error in deriving the
MF slopes for intermediate metallicity globular clusters. However, what is needed are
a range of models from the different groups in the redder HST colors (particularly F814W) where the theory is more successful in fitting
the observations.

As can be seen in Figures 8 and 9, there is an enormous
but systematic variation in the MF slopes as a function of radius. As shown here
(with the $F814W$ data) and in
Richer  et al. (2002) with the $F606W$ images, the MF slope in the outer field is very flat with
$\alpha = 0.1$. This is flatter than our result presented in Richer  et al. (2002) largely because of the way the incompleteness corrections were handled.
From the full matrix treatment (done here) it is clear that these corrections were somewhat over-estimated earlier. In the inner fields the slope is
actually inverted taking on negative values and decreasing from $-0.2$ in annulus 3 through to $-0.7$ in the inner annulus. This is a clear signature of extensive mass segregation which we discuss in some detail in \S 7.  

\section{Towards the Hydrogen-Burning Limit in M4}

We repeat here a statement made in Richer  et al. (2002), namely that
the least luminous MS stars observed in M4 in the outer field (where the data are deepest) are fainter than any other known subdwarf at the M4 metallicity (Leggett  et al.\ 1998; 
compare LHS 1742a and LHS 377 with Figures 1 and 2 in Richer  et al. 2002). We went on to conclude that a consequence of this was that
there should be correspondingly faint red subdwarfs in the field waiting to be discovered. It was thus of some interest when L\'{e}pine, Rich \& Shara (2003)
announced the discovery of a subdwarf with properties similar to the least
luminous stars found in M4.

This raises the natural question as to whether these stars are near the
hydrogen-buring limit for metal-poor objects. According to the Montalban
models the lower limit for hydrogen burning at this metallicity is $0.085 M_{\odot}$. The apparent $F555W$ magnitude for such a star in M4 is fainter than 35, while the apparent magnitude for stars only $0.005 M_{\odot}$ more massive is fully 4.5 magnitudes brighter.
This illustrates how extensively the mass-luminosity relation in the visible bands stretches out the cluster main sequence near the hydrogen-burning limit.
These same models, however, suggest that 
 deep observations in the near infra-red would be better suited than optical 
 observations for achieving the goal of reaching the end of the main sequence. We have in hand such data from the
Gemini Telescope and their reduction is currently in progress.

However, 
we can make some progress towards delineating the end point of hydrogen-burning in M4. For this purpose we will use the more recent deeper data taken in $F606W$ and $F814W$ from GO 8679. We note that in this data set the number of main sequence
stars between our saturation limit ($F606W = 19.4$, mass 0.56$M_{\odot}$)
and where we start to seriously run out of stars ($F606W = 27.8$, mass 0.095$M_{\odot}$) in our outer field
is 520. The adopted MF slope between 0.095 and 0.56$M_{\odot}$ is $0.1$ and if we extend this slope to the
hydrogen-burning limit at $0.085 M_{\odot}$ we predict 19 stars fainter than $F606W = 27.8$. 
A reasonable estimate of the error in the slope is $\pm 0.2$ given the
comparison between its measured values in $F555W$ and $F814W$ from the previous section. Using this
range in slope the expected numbers below $0.095 M_{\odot}$ run from a low of 15 to a high of 49. We
count only 6 such objects in our frames. The implication here is that if the MF continues
with the same slope all the way to the bottom of the main sequence
then there are at most a few tens of
main sequence stars that are awaiting discovery at the faint end in this M4 field. A good
test of this will be the Gemini data and whether it yields very red cluster
stars not seen in the
HST images.

\section{Michie-King Models for M4}

Our aim in constructing Michie-King models for M4 is to constrain its IMF from our data. The first task is to determine the distribution of mass within the cluster today and the second is to relate the results to plausible initial conditions. An important new constraint obtained from our work is the observed
number of white dwarfs, which we have argued (Hansen  et al. 2002) is a substantial fraction of the remnants formed in this cluster. A basic difficulty is that M4 has undoubtedly lost an important fraction of its initial complement of stars through dynamical evolution and that this loss is differential: low mass stars are preferentially lost. Consequently, the interpretation of the present-day state of M4 must be guided by dynamical theory, which, as we discuss later, has only a limited capacity to address the observations of real clusters.  

\subsection{Comments on the Michie-King Models}

Multi-mass Michie-King models \cite{GG79} are commonly used to convert {\it local} MFs determined within a field at some radial distance from the cluster center, into {\it global} MFs. These models, which we refer to as MMK models, are based on the assumption that each mass species can be described by a lowered Maxwell-Boltzman velocity distribution within a finite (tidal) radius. Moreover, the models are generally computed by setting the central squared velocity dispersion for each species to be inversely proportional to the species' mass. This assumption gives rise to mass segregation in the model even in the absence of strict thermal equilibrium (because the models have a finite size). The models may include  velocity anisotropy but only those with an isotropic velocity distribution will be considered below.

Dynamical models usually start with an assumed homogeneous mix of stars (e.g. by setting the velocity dispersion of each species in an MMK model to a constant value) and the calculations show that well developed mass-segregation arises only very late in the evolution, generally just before core-collapse (e.g., Chernoff and Weinberg 1990). Consequently the observation of significant mass segregation, particularly a core MF that is strongly biased toward high masses, is often taken to be a sign of incipient core-collapse or perhaps a post-core-collapse state (see, e.g., Lee  et al. 1991; Baumgardt and Makino 2003; hereinafter BM03). Consequently, one might be able to infer the present dynamical state of a cluster simply by comparing the degree to which the observed mass segregation is reproduced by an MMK model.             

 The utility of the MMK models is deceptive. Since they can involve a number of 
 largely unconstrained parameters, they are capable, through fine tuning,
 of fitting almost any data set and, 
 consequently, their predictive value is limited. In addition to the 
 questionable assumption of imposing a thermal equilibrium 
 condition in the standard MMK model, the applicability to 
 real clusters is further clouded by the likely 
 presence of unresolved binaries in the cluster core which may act as 
 an energy source that is unaccounted for in the MMK formalism. 

\subsection{Constructing the Input MMK Mass Function} 

A basic input into the MMK models is the cluster MF 
which should include the observed stars, the largely unseen 
remnant population and any brown dwarfs. 

In our MMK models, we start with a MF that is split into two power
laws with slopes of $\alpha_1$ for the low mass stars 
and $\alpha_2$ for the high mass ones. The MF is assumed to match at the break 
point. The intent of this
procedure is to mimic the shape of  an IMF
by converting the massive stars into remnants in a self-consistent way.
 In general, the break mass in this kind of formulation is a free parameter. With this construction, 
we obtain the following expression for the global ratio 
of high mass stars ($N_2$) to low mass stars ($N_1$): 
\begin{equation}
\frac{N_2}{N_1} = \left(\frac{1-\alpha_1}{\alpha_2 - 1}\right)\left[1-\left(\frac{m_2}{m_3}\right)^{\alpha_2 - 1}\right]  \left[ 1 - \left(\frac{m_1}{m_2}\right)^{1-\alpha_1}\right]^{-1}.
\end{equation}
In the above, $m_1,\  m_2$, and $m_3$ are the lower mass limit, the break mass, and the upper mass limit of our adopted MF. The value of $N_2$ is equal to the number of white dwarfs, since we assume that all stars in the upper mass range leave a remnant. 
Hence once we set the mass ranges, 
$m_2/m_3$, and $m_1/m_2$, in this equation, our knowlege of $\alpha_1$ and the
observational constraints on the ratio of $N_2/N_1$ from the data 
permit us to find a unique solution for $\alpha_2$.  

Turning first to the the low mass end, we note that 
there is considerable theoretical evidence that the MF at 
the half-mass radius is similar to the global MF (e.g., BM03), 
and since the half-mass radius of 
M4 falls very near to  our outer field, we take the MF in this region 
to be the global MF for the main sequence stars. Hence, in accord 
 with the results reported in \S 5, 
 we adopt $\alpha_1=0.1$ for the present day MF 
 over the mass range of $ 0.08 \le m \le 0.8$. 
the Montalban models used here
 place the present day turnoff mass close to $m_{to} = 0.8 M_{\odot}$

The lower
 mass limit is essentially the termination mass for the hydrogen-burning 
 main sequence and, therefore, these models do not include any brown dwarfs.
 If sub-stellar masses follow the observed mass function, 
 they can have little influence on the dynamical structure and 
 thus neglecting their existence will not affect the
 comparison of the model results to the observations reported here. In the
absence of any clear evidence to the contrary, we prefer to set the break
point in the adopted MF at the main sequence 
turn-off mass. We have investigated models 
with different break points but, as shown below, we obtain good 
agreement with the data for a model with the break point
set at the turn-off. This is ad hoc and simply a matter of convenience.
One consequence of the chosen global mass intervals, is if that $\alpha_1 +
 \alpha_2\,=\,2$, which, as we argue below, may be approximately the 
case in M4, then the global ratio $N_2/N_1\,=\,1$, 
implying that we might expect about  half the total 
stellar population today to be white dwarfs.

 The stars that once were above the turn-off, in the decade 
 $0.8 \le m \le 8.0$,
 are assumed to be white dwarfs today.  
No neutron stars are included in the 
 adopted cluster MF's  discussed here even though M4 is 
 known to have at least one millisecond pulsar (Lyne  et al. 1988).
 This omission 
could have some effect on the very central density because the heavy neutron 
stars are expected, as a class, to be very centrally concentrated. 
The bottom heavy MFs used below would not yield very many neutron stars 
in the first place and many
would likely recoil out of the cluster upon formation. We 
do not believe that the absence of 
neutron stars affects the modeling reported here.
 
In the outer field with our deepest data in $F606W$ and $F814W$ 
the number of main sequence stars (corrected for incompleteness)
between masses 0.095 ($F606W = 27.8$) and 0.56$M_{\odot}$ ($F606W = 19.4$)
 is 520. Extending this up to the turnoff
mass at $0.8 M_{\odot}$ with slope $0.1$ as derived and 
down to the hydrogen-burning limit at $0.08 M_{\odot}$ suggests a 
total current number of 786 (this is $N_1$). In our deep data 
in the outer field  the number of white dwarfs observed 
to the 50\% completeness limit is 290. We refer to this as $N_v$,
rather than $N_2$ because these are, of course,  not 
all the cluster white dwarfs in
our field. There is certainly a population fainter than 
our limiting magnitude that
are He white dwarfs or extremely low luminosity H-rich ones.

To  account for the component of cluster white dwarfs 
that have not been detected, we proceed as follows. 
At $F606W = 29.1$ we reach the 50\% completeness limit 
in our photometry for white dwarfs. Below this
the numbers become unreliable. For a white dwarf 
with a typical mass ($0.6M_{\odot}$), the
age at this luminosity is 9.5 Gyr. 
If we assume a cluster age of 12.1 Gyr as
derived in Hansen  et al. (2004), the main sequence lifetime of 
the progenitors of these
white dwarfs was 2.6 Gyr. The mass of a star with this hydrogen-burning
lifetime is $1.3M_{\odot}$ according to the models of 
Hurley  et al. (2000). If we then use this mass as $m_3$ 
in equation 1 we derive from the ratio of $N_v/N_1$ that
$\alpha_2\, =\, 2.3$ - essentially the Salpeter slope and considerably steeper
than the low mass slope in M4. Taking into account the uncertainties in the
star counts, $\alpha_2$ could be as low as 1.5 or as high as 2.5. 
In the MMK model presented below, we have used a slope of 
$\alpha_2\,=\,2.3$ applied to the entire range $0.8-8.0$. 

The possible connection between the present-day MF and the IMF is 
discussed in \S 7.4. Here we note that the visible
white dwarfs, most of which were produced in the distant past when 
the cluster mass was likely much higher than it is today, have a mass that 
is quite close to the mean mass of today's main-sequence stars.
Consequently, both species will have suffered quite similar 
evaporative losses over the history of the cluster. In other words, 
we expect that the ratio of today's visible white dwarfs to 
today's visible main-sequence stars,  $N_v/N_1$, has  not 
changed very much and, therefore, the indicated value of $\alpha_2$ is
applicable to the epoch when the white dwarfs formed. 

It is of interest that the derived slope is so 
close to the Salpeter value. Given the ubiquity of this function in Pop I 
environments (see, e.g., Elmegreen 2000) it perhaps is not surprising that
it shows up in Pop II as well. Moreover, the flattening of the IMF 
at some low mass is also ubiquitous and may also be a feature of 
Pop II (Reid and Gizis 1999, Zoccali  et al. 2000, Kroupa, 2001, Paresce 
and De Marchi 2000).

\subsection{Constructing the Michie-King Model for M4}

In an MMK model the continuous distribution of stellar masses is 
replaced by a series of discrete mass bins.  Here, there are 12
logarithmically defined mass bins assigned to the low mass 
stars between 0.08 and 0.8 solar
masses, and 6 to the high mass stars between 0.8 and 8.0, which include
a bin between 0.8 and 1.2 to allow an easy comparison between the model
and the visible white dwarfs. 
The stars in the high mass bins were converted to white dwarfs 
using Wood's (1992) exponential initial-final mass relationship 
for a turn-off mass of $M_{TO}\,=\,0.8M_{\odot}$:
\begin{equation}
 M_{wd}\,=\,0.510M_{\odot}e^{0.095M_{ms}}.
\end{equation}
The mean mass of the model white dwarfs is $<m>_{wd} = 0.56$.

Once the MF is fixed, the MMK models are a single-parameter family 
characterized by a normalized central potential $W_0$. In addition,
Each MMK model has two scale parameters that may be determined from observations 
of the surface brightness profile (i.e., the radial distribution of the stars): the
cluster scale  radius ($r_s$, a model parameter that approximately 
corresponds to the core radius, the  distance
 where the surface density falls to about 1/2 the central value), 
and the projected central density. For our purposes, there is no 
need to independently determine that scale parameter, 
which would, in any case,  be difficult with the data set we 
are using to define the cluster radial profile. 

Since each mass class 
will have a different core radius, some
 care must be taken to ensure that the model mass component being fitted to the 
observations matches the  quantity used to define the observed profile: star counts
 in some magnitude range or integrated light in some bandpass. Here we use 
surface brightness data compiled by Carlton Pryor, which includes his original 
 counts of bright stars ($V \leq 16$) within 
the  inner $30\arcmin$ of M4, supplemented  
by an integrated light profile from a Sky Survey photographic plate, 
PS-7586E, that covers
 the outer part of the cluster.
The observed profile is that defined mainly by giants and 
upper main-sequence stars. 
In our models, we have a mass class at $0.69 M_{\odot}$, 
which we take to  represent the upper 
main sequence stars (including the giants), and we use the calculated radial 
distribution of these stars to compare with the observations.

For a given choice of $r_s$, we determine the best fit by spline-interpolating
on the model curve at the radius of each of the observed data points and 
minimizing the deviation.  We calculate $\chi^2$ for the fit as the 
 sum of $(O-C)^2/\sigma^2$, where $\sigma$ is the quoted error of the data point.
The star counts were fitted independently
of the integrated light data at the outer radii. The difference in observational
units for these two representations of the cluster profile corresponds to 
different density normalizations for the two data sets and, in the absence of
a pre-determined connecting parameter, they must be treated 
separately to bring them onto our common model curve. 
We constructed  models with $W_0$ between 8 and 10, and for each, varied $r_c$ 
systematically to find the minimum $\chi^2$. The star counts were used as the
determining data set; the integrated light points were used as a check.   
 The best fit was obtained with $W_0 = 8.5$ but we note that the correlation
 between $W_0$ and $r_s$ is such that almost indistinguishable 
fits could be obtained 
for $W_0$ values of 8.5 and 9.5 with slightly larger or smaller values of $r_s$. 
Our preferred fit is shown in Figure 10. The fit
to the inner points (circles) has $\chi^2 = 1.28$, and the outer points (squares)
 have  $\chi^2 = 0.92$. The scale radius for this 
model is  $r_s = 55\arcsec$, and the {\it core} radius of the fitted mass-species
is $r_c = 53\arcsec$, in
good agreement with the Trager  et al. (1993) value derived from single-mass
 King models fitted to a compilation of surface-brightness data \cite{TKD95}.

 The calculated mass functions for
 our four annuli are plotted in Fig.\,11 to show how mass segregation
 distorts the input global mass function in our observed fields. 
This plot shows the predicted number of stars per unit solar mass, per 
square parsec at the cluster, at the median radius of the annulus that
includes each of our four fields. It is 
evident that if we adopt a global power law for an  MMK model, then 
the calculated mass functions at any particular radius in the model 
cannot be a power law. The deviation will be 
particularly marked close to the cluster center, i.e., 
in the inner two fields, as we see plainly in Figure 11.  
However, we note that the model output in field 4, which is near the
cluster half-mass radius, does display the global slope ($\alpha_1\,
=\,0.1$), apart from the very highest masses. For the other fields, the
power-law slopes discussed in \S 5 are crudely consistent with the trends
observed in Fig. 11. 

To connect this result back to the observed cluster luminosity functions, we
proceeded as follows:

\begin {enumerate}

\item 
We used a model mass-luminosity relationship (Montalban  et al. 2000)
to convert the model mass to luminosity (magnitude); in other words 
to change the model  mass function (as a function of radius) 
to a luminosity function (as a function of radius).

\item 
We spline-interpolated along the model luminosity 
function at each magnitude bin in the
data set to get the number of stars at each luminosity.

\item 
We fit the model output to the observed luminosity function and compared
with the data.

\end {enumerate}

The result of this is shown in Figure 12 for $F555W$ and Figure 13 for $F814W$. While the $\chi^2$ are not particularly good for $F555W$, they are acceptable for the $F814W$ luminosity functions.
This demonstrates that we have achieved reasonable concordance between
the presently observed M4 luminosity function and the predictions of our MMK model.

\subsection{Mass Loss from M4 and Dynamical Theory: Towards the M4 IMF}

 The relatively shallow MF currently observed in M4 at low mass cannot
 be the true IMF because the cluster as a whole has suffered a 
 substantial loss of its initial complement of stars 
through evaporation (e.g. Chernoff and Weinberg (1990; hereinafter CW90), \cite{DJ93}, BM03)),
 tidal stripping and disk-shocking (Vesperini \& Heggie 1997). The mass loss
 due to stellar evolution is an important dynamical driver but it is the differential 
losses between mass species that play a critical role in the interpretation of the data. 
 The MMK models discussed above do not include such dynamical evolution and 
 are strictly meant to allow a comparison with the observed state of the cluster.  
 Nevertheless, we assert that the qualitative features 
 observed today, namely, a relatively flat MF at low masses 
 and a remnant population that can be derived from a much steeper 
 slope at high masses, are expected to be present in the IMF 
 but with somewhat different quantitative values. To support our assertion, we will
briefly review the present state-of-the art in the relevant aspects of the 
theory of cluster evolution.

The pioneering survey of the evolution of globular clusters in the Galaxy was done by CW90 using Fokker-Planck theory. Their work has been updated with an improved theoretical treatment by Takahashi and Portegies Zwart (2000 hereafter TPZ00). The basic results are mostly unchanged: clusters with top-heavy IMFs (those with most of the initial mass in rapidly evolving high-mass stars) tend to dissolve and those with bottom-heavy IMFs tend to undergo a core-collapse. The time-scales for either of these end points depends on the initial mass of the cluster, its intial central concentration, and its location in the Galaxy. The Fokker-Planck  models show the expected preferential depletion of low-mass stars, which decreases the slope $\alpha$ of the MF at the low mass end and are the stars we observe today on the cluster lower main sequence. One important result from CW90 (their Table 6) is that changes in the MF slope are fairly small  ($\Delta \alpha \simeq 0.1 - 0.3$). A second relevant point is that clusters close to core-collapse generally exhibit strong mass segregation and will exhibit a top-heavy MF in the cluster core. A third point is that clusters at their end points generally have lost between 25\% and 75\% of their initial mass (TPZ00). Both the CW90 and the TPZ00 models were designed to facilitate a survey of generic properties and thus do not represent any particular cluster. They assume that the IMF is a single power law between $15 M_{\odot}$ and $0.4 M_{\odot}$. The artificial truncation of the low-mass end of the IMF will have some quantitative effect on the evolution of bottom-heavy models but mainly it limits the comparison of these models with observational data, which now cover most of the mass decade between the turn-off at about $0.8M_{\odot}$ and the hydrogen burning limit at $0.08M_{\odot}$.   

In parallel with the advances in Fokker-Planck theory, N-body studies of cluster dynamics has been greatly advanced through the development of the special purpose GRAPE machines (e.g., Makino  et al.  1997). Recently, BM03 have presented the results of their extensive survey of the evolution of star clusters in the tidal field of the Galaxy. In a departure from most previous studies, they have adopted the Kroupa (2001) IMF for their work. This IMF is represented by two power laws: $\alpha_2 = 2.3$ for the high-mass end and $\alpha_1 = 1.3$ at low masses, with the break point at $0.5 M_{\odot}$. These simulations included a variety of initial conditions and each model was calculated with a range of N, allowing the authors to quantify the dependence of their results on N (see Heggie 2000 and references therein for a discussion of this issue). The models were followed through core collapse to dissolution (defined as that time when 95\% of the intial mass has been lost). As in the Fokker-Planck models, differential mass-loss flattens the initial MF at a slow rate until just before final disruption, a point marked by very strong mass segregation. Of particular interest is the apparent evolution of the ``Kroupa kink" in the MF, which is gradually washed out and therefore, would be rather hard to detect in real data.    

BM03 have applied their results to an interpretation of M4 but the theory evidently fails to reproduce the observed characteristics of M4: the observed MF is far too shallow. There are at least three plausible reasons for this failure: (1) The assumed IMF is inappropriate. Paresce and De Marchi (2000) show that a log-normal MF appears to fit most of the deep globular cluster data now available and suggest this reflects the IMF. A shallower slope ($\alpha = 0.5$) than the Kroupa value has been suggested by Reid and Gizis (1999) for the field Pop II IMF, but their value is poorly constrained. A change to a shallower slope would not likely have much affect on the BM03 results because their low-mass IMF already has the mass biased toward the high mass end. Consequently, we might expect the rate of change of the mass slope to be qualtitatively similar to the published results. (2) M4 was assumed to be on a high eccentricty orbit ($\epsilon = 0.8$, Dinescu  et al. 1999) moving in a logarithmic potential, appropriate to the dark matter halo only. The orbit may not be as eccentric as indicated and M4 may well be affected by tidal shocks in moving into the disk and bulge of the Galaxy. A less eccentric orbit would slow the dynamic evolution whereas tidal shocks would speed it up; the net effect is unknown. (3) The simulations did not start with a primordial binary population and any binaries that formed in the course of dynamical evolution were treated as inert. As the primordial binaries harden, they release energy that will delay the onset of core-collapse (Gao  et al. 1991) and binaries that form during collapse will help power a post-core collapse expansion phase.  

In light of the preceeding remarks, the N-body models, like the Fokker-Planck models, are best regarded as indicative of trends. Overall, the general agreement among studies using different approaches, which, besides the above, include semi-anaytical calculations (Johnstone 1993) and updated Monte-Carlo simulations (Joshi  et al. 2001), is comforting. The qualititative evolution of the cluster IMF in all studies is well illustrated in Figure 7 of BM03.   

The conclusions to be drawn from the observations are that M4 is in an advanced dynamical state. The evidence is the marked mass segregation and the sharply inverted mass spectrum in the central regions: it must be approaching a state of core collapse (CW90, BM03). The MMK model that describes the cluster today has 52\% of the mass in the form of remnants and this high fraction is also an indication of extreme dynamical evolution (BM03). In view of the rapidity of the MF evolution toward the end of the cluster lifetime (BM03), it is rather important to try to pin down the dynamical age. From the surveys of CW90 and TPZ03, it is likely that M4 started life in a fairly concentrated state (like the $W_0 = 7$ models in the surveys), since otherwise it would have disappeared by now given its location and orbit. Such models will typically have lost 35-70\% of their initial mass (depending on the exact mass spectrum) upon entering core collapse (TPZ03). Comparing the BM03 results for their families I and II, which differ only in the adopted initial concentration ($W_0=5.0$ and $W_0=7.0$ respectively), we note that the initially more concentrated models enter core-collapse sooner and that their time to dissolution is about twice the core-collapse time. Applying this precept to M4, assuming it is about to go into core-collapse, the cluster would have a dynamical age of $T/T_{diss} \simeq 0.5$ in the notation of BM03. Removing the quasi-impulsive mass lost due to the nuclear evolution of the stars (about 30\% of the initial mass for a Salpeter-like IMF), the dynamically induced mass loss is approximately linear with time. Hence this argument leads to a present day mass that is about $1/3$ of its initial value. In this case the evolution of the MF slope for the 
present main sequence stars would still be relatively modest - typically a 
few tenths in the BM03 N-body models, which is similar to the predictions 
of the Fokker-Planck models (CW90). Hence the IMF of the low mass stars was likely quite close to being flat to begin with, say $\alpha_1 \simeq 0.4$. 
The location 
of the break mass is not discernable in our data but it is probably somewhat smaller than the turn-off mass that was adopted here. The remnant white dwarf population has a mean mass that doesn't differ too much from the mean mass of the present day main sequence stars ($0.56$ and $0.44 M_{\odot}$ respectively in our MMK model) and to within a factor of order unity (CW90, Fig. 33a), both populations suffer a similar dynamically driven loss of stars. This last point alone, together with our observations, provides quite a compelling case that a steep IMF applies to the white dwarf progenitors independent of any detailed evolutionary model.

\section{Summary and Discussion}

The principal results from this multi-field photometric study of the main sequence of the globular
cluster M4 are summarized below.

\noindent $\bullet$ The cluster main sequence has been traced to the limit of our data at 
 $F555W \simeq 27.5,\ F555W - F814W \simeq 3.8$. Stars this faint are
less massive than $0.1 M_{\odot}$ according to the current generation of models.

\noindent $\bullet$ A simple analysis of 
  the observed scatter about the cluster main sequence suggests a rather small
overall binary fraction with 
 weak evidence for a radially dependent binary fraction, $f_b$. For stars with $M_{F555W}\ga 7.5$ 
 ($M \la 0.6 M_{\odot}$),   
 we find that $f_b \simeq 2\%$ 
 in the cluster core but this falls to the $1\%$ range outside. Our data thus
mildly support the suggestion of \cite{CF96}
 that M4 has a high central binary fraction which could be supplying the
 energy needed to prevent core-collapse in the cluster.  

\noindent $\bullet$ The main sequence 
 luminosity function differs significantly from field to field, and the
 data as a whole display a striking drop in the star counts at $F555W\simeq 24.5$.
  
\noindent $\bullet$ The local cluster mass functions
 determined with the Montalban  et al. (2000) isochrones are quite flat
in each field. The abrupt drop in the main sequence
 density noted above is due to the slope of the mass-luminosity relationship
 at low masses.
 The goal of reaching the end of the hydrogen-burning main sequence 
 was not achieved here nor 
 has it been achieved in much deeper data (Richer  et al. 2002) nor
in NGC~6397, the only other cluster for which
 such a goal is within reach \cite{CPK96}. The 
 existence of a minimum stellar mass is a fundamental theoretical result 
 that can, in principle, be confirmed. Although our data appear to reach to within $\simeq 0.01 M_{\odot}$ of the theoretical limit, 
 the difficulty 
 is that the mass-luminosity relationship becomes very steep as the limit is
 approached. 
 The main sequence observed in the optical bands becomes
 very sparse and extends to extremely red colors: $(V-I) \ga 5$ for the
 Montalban  et al. (2000) models. 
 These same calculations, however, suggest that 
 deep observations in the near infra-red would be better suited than optical 
 observations for achieving this goal.  While there is no direct evidence that we have 
 reached the end of the hydrogen-burning main sequence, extending the cluster
mass function to this limit suggests that there are rather
few cluster stars fainter than our limiting magnitude.

\noindent $\bullet$ The observed variation in the cluster mass function
 from field to field is
 consistent with the segregation found in isotropic, multi-mass Michie-King
 models provided the present-day global mass function
 has a slope near the observed value of $\alpha = 0.1$. However, an accurate definition of the  global cluster mass function is still not 
 possible with the current data set. There are two reasons for this. The first 
 has to do with the conversion of luminosity functions
  to mass functions through
 stellar models which provide the necessary mass-luminosity relationship. The
 models used here do not provide a reasonable fit to the 
 extended main sequences revealed by HST observations of M4.
  Clearly further improvements are needed to bolster confidence in the local mass
 functions derived directly from the data. However, there is a suggestion that
this may not be such a serious problem as the slopes derived independently from
the $F555W$ and $F814W$ data are in excellent agreement.
 The second reason is that mass segregation 
 occurs and, in the absence of complete radial coverage of the cluster,
 must be modeled to interpret the local mass functions. 
 Multi-mass Michie-King models can be used for this purpose, as was done 
 here. However these models have little predictive power and their
 applicability {\it as physical models} to globular clusters, particularly
 objects like M4, which appear to require an internal energy source to avoid
 core collapse, is questionable. Fortunately, the observational problem of 
 determining the global 
 luminosity function can be addressed simply by observing a complete
 radial cross-section of the cluster.
 HST observations are certainly needed in the core, but
 deep ground-based data may suffice for the halo regions. In principle, the
 present-day mass function can be related to the initial mass 
 function through dynamical modeling, an ever improving art \cite{MH96}.

An important motivation for further dynamical modeling of M4 is
 to provide some insight into the white
 dwarf remnant population now residing in the cluster. Recent steps in this 
 direction are described in the work of Shara \& Hurley (2002).
 As noted in Hansen  et al. (2002, 2004),
 the cluster white dwarf luminosity function
 provides a new, independent age estimator. As this measurement becomes more refined, the dynamical evolution of this remnant population
 becomes an increasingly important issue to be understood.

\begin{appendix}
\section{Details of the Data Reduction}

\begin{enumerate}

  \item The data were preprocessed according to the recipe given in the
  ALLFRAME cookbook (Stetson 1994, Turner 1997). To summarize, vignetted regions were 
  masked out. The images were multiplied by 4 (as this allows for short integers to be used while still sampling below the read noise) and saved as
  reals. They were then multiplied by an illumination
  map which adjusted for how each pixel was illuminated by the optics, and
  split into pc, wf2, wf3, wf4 fields.  

  \item Using a C-shell script, the `sky' was measured on each image and plots made of sky-level versus
  sun angle (see Figure 1). This was done to investigate whether
  any special strategy should be employed when combining the data
  frames.

  \item  For each individual frame, we used DAOFIND with a
  4-$\sigma$ finding threshold to generate a star list. This list was
  cleaned by rejecting objects found near the edges, and the stars were
  photometered using PHOT and ALLSTAR. Library PSFs provided by Stetson were
  used.

  \item \label{matchall} During the data acquisition, the individual frames had been dithered by
  small pixel amounts. An initial guess was made that
  the frames were aligned, and this guess was refined by DAOMASTER to
  produce 6-term (translation and rotation) transformations between the 
  frames (the MCH file).
  Note that {\it all} of the data, both old and new, for a given CCD was matched at the
  same time so that all frames were tied to the chosen reference frame.

  When matching frames, the DAOMASTER program determines a
  magnitude offset between the frames that is later used by the
  MONTAGE2 program to scale frames before combining. We set this
  parameter to zero. We did this as: (1) there is no variable atmospheric
  extinction in space; and (2) we had different filters in our matching
  list, which led to artificially large values of this parameter. A few
images had slightly different exposure times and these were   
appropriately scaled. 

  \item \label{montage} The MCH file from step \ref{matchall} provides geometric
  transformations that map the nth image in the MCH file onto the
  coordinate system of the first image in the MCH file (corollary: the
  first line in a MCH file is an identity transformation).

  The MONTAGE2 program - which was designed primarily for generating finding charts from partially overlapping frames - combines the images in a MCH file by transforming
  and then taking a median.  Instead of using MONTAGE2 to combine all
  of the images in our MCH file, we split our MCH file into individual
  lines (the nth line of the MCH file represents the geometric
  transformation between the nth image and the first image) and used
  the individual lines as input to the MONTAGE2 program. This resulted
  in producing (for each image in the MCH file) a new image that was
  on the coordinate system of the first image in the MCH file.

  In summary, we transformed all the images for a given CCD to a
  common coordinate system, and scaled the few images which had slight
  differences in exposure times.

  \item \label{combine} We combined those registered frames which had the same epoch
  and same filter. Four images (oldv, oldi, newv, and newi) were
  produced for annulus 4, whereas three images were produced (oldv, oldi,
  and newi) for the inner 3 annuli.

  The registered frames were combined using IRAF's IMCOMBINE task to
  reject the brightest $n$ pixels, and averages (not medians) were
constructed of the remaining
  pixels.  This method of combining was preferred as it provided
  us with higher S/N images; the noise in the average of two
  frames will be similar to the noise in the median of three frames (Stetson 1994). 
  Hence, provided we reject less than 1/3 of
  the pixels, the noise in the high-pixel rejected averaged frame will
  be less than that in a median of all the data. In practice this technique
allowed us to use 93\% of the $F606W$ frames in the long exposures in annulus 4 and
94\% of the $F814W$ images in constructing the final images. These advantages
were much less spectacular for the inner annuli where the number of individual images
was much less. 

  By rejecting high valued pixels we reject the cosmic ray events
  (CREs) which are present in the individual frames. The HST manual
  states that 20,000 pixels are affected by CREs in an 1800s
  exposure. This gives a CRE rate of $1.736 \times 10^{-5}$ CRE pixel$^{-1}$ sec$^{-1}$. The mean
  number of CREs we expect in a pixel is the exposure time multiplied
  by $1.736 \times 10^{-5}$. As the CREs follow a Poisson distribution,
  the standard error of the mean is the square root of the mean. To
  ensure that CREs are effectively rejected, the number of pixels we
  reject is 5 standard deviations higher than the expected number. By visual
inspection, this process was quite effective in removing cosmic ray events 
on
our images. 

  \item \label{mkcoo} A finding list was generated for the {\it combined} frames.
  Great care was taken in doing this by proceeding as follows:

  \begin{enumerate}

    \item A 2.5 sigma find was made on the $V$ and $I$ frames.
    The resulting files were matched and stars in common to both frames
    were used in the initial candidate list.

    \item \label{initial} Using ALLSTAR, stars from the candidate list
    were subtracted from the deep $V$ and deep $I$ images.

    \item \label{findv} A median filter was used to smooth the unsubtracted
    $V$ frame, and this median filter frame was subtracted from the $V$ 
    frame and the $V$
    frame without stars. These two images were displayed, stars from
    step \ref{initial} circled, and then the two images were blinked.
    With careful inspection, missed stars were added into the finding
    list, and spurious detections (e.g. lumps on diffraction spikes) were
    rejected. As a point of interest approximately 8 person-hours were
expended in constructing the finding list for each chip.

    \item the procedure in \ref{findv} was then repeated for the $I$
    frame (in order to find red objects that might have been missed on the bluer $V$ frame), with the difference that the most recent coordinate list was
    used.

  \end{enumerate}

  \item \label{als1} The coordinate files from step \ref{mkcoo} were
  used along with Stetson's PSFs to photometer the combined frames with
  the ALLSTAR program. Note that ALLSTAR has a recentering option  which
  we enabled so as to redetermine the stellar center when performing
  the PSF fit.

  \item \label{padding} If ALLSTAR was
  unable to converge on the magnitude of a star,
  that star does not appear in the output ALLSTAR list. A
  consequence of this is that the ALLSTAR files from step \ref{als1}
  contained different subsets of the stars from the coordinate files
  used as finding lists. To simplify analysis, we `padded' the
  ALLSTAR files with null data so they had a 1-to-1 correspondence
  with the input coordinate file (note that ALLSTAR preserves the id
  number used in the input coordinate file). 

  \item Recall that in step \ref{matchall} we matched stars on {\it
  all} frames (both new and old). As the M4 fields are dominated by
  cluster stars, the transformations were dominated by cluster stars
  (indeed, as the matching radius was decreased, the non-cluster stars
  were preferentially rejected).  Hence, the frames were registered
  using primarily cluster stars. A consequence of this is that non cluster
  stars appear to move when frames from the two epochs are blinked. As
  we allowed ALLSTAR to redetermine the stellar coordinates, on a plot
  of $(x_{new}-x_{old})$ vs. $(y_{new}-y_{old})$ we see the cluster
  stars grouped around the origin, with a secondary group of non-cluster
  stars displaced from this.

  \item To improve the transformations between the individual frames,
  we generated a list of stars likely to be cluster members by
  isolating the clump of stars near the origin on the
  $(x_{new}-x_{old})$ vs.\ $(y_{new}-y_{old})$ diagram. A CMD of the
  selected stars indicated that the selection was successful.  The list
  of candidate cluster stars is on the coordinate system of the
  transformed frames.  We transformed this coordinate list back onto
  the system of each of the non-registered frames by inverting the
  transformation from the MCH file.

  \item Using the list of cluster stars, the individual
  (non-transformed) frames were reduced using the ALLSTAR program with
  the recenter option enabled.  These `cluster only' ALLSTAR files,
  which have only cluster candidates, were matched with DAOMASTER, using
  a 20 term transformation equation. By experimentation, we found that
  the 20 term transformation reduced systematic errors. Plots were made
  of $dx$ and $dy$ for the cluster stars, and streaming motions (indicative of a poor fit leading to large residuals) especially
near the edges of the chips were
  seen when a 6 term transformation was used. As prescribed by Stetson, the DAOMASTER matching
  radius was reduced until it was similar in magnitude to the rms error
  in the transformation. The 20 term transformation between the frames (based
  on cluster stars only) was saved to a new MCH file.

  \item As in step \ref{matchall}, we edited the magnitude offset
  values in the new MCH file.

  \item As in step \ref{montage}, we used the MCH file to produced
  transformed images. Note however that this time we used the MONTAGE2
  program to expand the frames by a factor of 2 (pc CCD) or 3 (wf
  CCDs). The expanded, registerd images were then combined using the
  same procedure as described in step \ref{combine}.

  By expanding the frames, we found that the measured radial dispersion in the cluster
  proper motion was reduced from 3.9 to 2.5 mas/year. The intrinsic dispersion amongst the M4 stars
  is estimated to be about 0.5 mas/year, so the measured dispersion is representative
  of the errors associated with our technique.  In effect, by expanding
  the frames we implemented a form of `drizzling' (Fruchter \& Hook 1997).

  \item A typical background value was added back to the combined
  frames (MONTAGE2 subtracts off the background when transforming the
  frames). The DAOPHOT.opt and ALLSTAR.opt files were modified to take
  account of the different parameters (gain, fwhm) on the expanded
  frame, and the master coordinate list was expanded so as to match
  the expanded frames.

  \item \label{psfs} Stetson's library PSFs were not appropriate for the
  expanded frames, so new PSFs for these frames were built following
  standard techniques outlined in the DAOPHOT manual.

  \item The expanded frames were photometered using ALLSTAR along with
  the expanded coordinate list and the PSFs built in step \ref{psfs}.
  The ALLSTAR files generated were `padded' as outlined in step
  \ref{padding}.

  \item All objects in the master coordinate (.coo) list were visually 
  classified (as to stellar or extended, isolated or possibly contaminated), and this clasification was saved in the .coo file although in the final analysis no use was made of this classification.
  
  \item As a check, a plot was made of the
  measured motions of the cluster stars on the sky. If the the old and new
  epoch are badly aligned, a `streaming motion' is seen near the edges of the chips. Such a motion
  was noticed when a 6 term transformation was used to match the
  data, but this motion disappeared and the motions appeared
  random when the full 20 terms were used.

\end{enumerate}

At this stage, the proper motion of a star (relative to the cluster) can be
determined by examining its shift in $x$ and $y$. Cluster stars can be
isolated in the data as they will have small proper motions.
An important point to note here is that in the outer annulus we were matching
long exposures in $F606W$ (127.4 ksec) and $F814W$ (192.4 ksec) with much shorter
exposures in $F555W$ (31.5 ksec) and $F814W$ (7.2 ksec). The question thus arises
as to how we could measure positions in the earlier epoch data for the faintest stars where the s/n is obviously poor. For a star at $F606W = 29.0$ the expected
s/n (from the HST ETC) in our deep $F606W$ frames is 6.7 while a star at this
magnitude on the shorter exposure $F555W$ frames has s/n = 2.5. This is just about
what we measured on our frames. The reason we were able to measure positions
of such faint objects in the first epoch are severalfold.

First, we applied the
finding list from the deep frames to the shallower frames. If the background noise in an image is Gaussian, then a
$2.5\sigma$ positive deviation will occur in about 6 pixels out of every 1000.
If, in a $750 \times 750$ image (neglecting the vignetted areas at the low-x and low-y
sides of the WFPC2 CCDs), we were to mark all of the $2.5\sigma$ peaks as
detected astronomical objects, we would expect more than 3,000 false
detections.  However, if we consider only the area within 0.5 pixels radius
of an object confidently detected in the long second-epoch exposure, we
expect a probability $\sim \pi \times 0.5^2 \times {6\over1000} \approx
0.005$ of finding a positive 2.5$\sigma$ deviation which is purely the result
of random noise in the first-epoch image.  That is to say, we expect of
order 5 false cross-identifications for every 1000 correct re-detections.
Even at $1.8\sigma$, presumed re-identifications will be correct 19 times out of 20.
For this reason, the knowledge that an actual astronomical object is present
somewhere nearby based upon the long-exposure second-epoch images allows us
to be confident that most of the claimed re-detections on the short-exposure
first epoch images correspond to true re-detections.  The first-epoch
astrometric positions, then, while poorer than those of the second epoch,
are nevertheless good enough to distinguish stars that are moving with the
cluster from stars and galaxies that are not.

Secondly, the proper motions of the
M4 stars are
actually quite large (about 1 HST pixel with respect to an extragalactic background over the 
6 year time baseline (Kalirai et al. 2003)).
While the lower $S/N$ image would not produce good enough photometry,
the $S/N$ is sufficient to give the centroid, which is crucial for
astrometry.
In Figure 2 we illustrated the quality of the proper motion
separation between cluster and field from the long and short images in the outer field.
Some of the brightest stars do not
exhibit clean separation due to their near saturation. However, what is clear from this diagram is that most of the
field objects can be easily eliminated down to very faint magnitudes ($F606W = 29.0$) by making 
the generous proper motion cut within a total value of 0.5 pixel ($\sim 8$ mas/yr) of that of 
the mean cluster motion over the 6 year baseline
of the observations. 
Since we did not push the data to its
limits in this contribution, for our present purposes, we tightened this constraint to
0.3 pixels which had the effect of reducing the scatter in the resulting CMDs.

\section{The Inner Halo Stars}

\subsection{The Inner Halo Main Sequence}

 The line of sight to M4 passes through the Galactic inner halo, or, perhaps more
 correctly, the inner Population II
 spheroid at a tangent radius of $R_t = R_o sin\phi$, 
 where $\phi = 18.3\arcdeg$ is the angle between the cluster line of sight
 and the Galactic center. 

 On a CMD, the stars in the inner halo will appear as a broadened 
 sequence because
 they are strung out along the line of sight and may differ in age and 
 metallicity.  However,  for a sufficiently steep density law,  the stars 
 will be well concentrated in a locus determined by 
 the distance modulus of the maximum apparent stellar density. 
 This distance, $d_{max}$, 
 is slightly beyond the tangent point distance, $d(R_t) = R_o cos\phi$,
 because of the increased volume 
 sampled by the solid angle of the field of view. 
 For the spatial density law, $\rho 
 \sim R^{-3.5}$ (Zinn 1985) , 
 the volume correction factor is $f = d_{max}/d(R_t) = 1.061$. 
 Therefore, the inner halo main sequence will appear below that of M4 on a CMD 
 due to the difference in distance moduli, $\Delta \mu$, where,
\begin{equation}
 \Delta \mu = 5log(f\,R_o\,cos\phi/d_c) = 5log(R_o/d_c) + 0.016,
\end{equation}  
 and $d_c$ is the distance to M4.\footnote{
 Although a simple power law with an exponent, $n=3.5$, 
 characteristic of the stellar halo, 
 may be an over-simplification of the true 
 density distribution along the line of sight 
 (see the discussion in Minniti 1995),
 the volume correction factor is quite insensitive to $n$:
 $f=1.053$ and $f=1.071$ for $n=3.0$ and $n=4.0$, respectively. This leads to
 less than 1\% uncertainty in the distance to the stellar density maximum, a
 negligible amount compared to other sources of error.} 
 This equation assumes that there is no
 additional reddening beyond M4 because the cluster 
 is located $\simeq 500$ pc above the Galactic plane. 
 Hence a measurement of the displacement, $\Delta V$,
 between the cluster and inner halo stars on
 the CMD will determine the ratio $R_o/d_c$ with an accuracy which may be much
 better than that attained for either $R_o$ or $d_c$ separately.

 Inspection of the inner halo CMDs in Fig 3. (right hand panels) clearly 
 reveals the presence of such
 a sequence. This inner halo main sequence is particularly well defined brighter
than $V = 23$, while fainter it may be contaminated by some redder
population which may be from the disk or thick disk although its kinematics are indistinguishable from that of the inner halo. The small color spread of the inner halo CMD
suggests that it is meaningful
 to assign an average age and metallicity to this stellar population.

 \subsection{The M4 - Galactic Bulge Relative Distance}

For a fixed age,  clusters at higher 
 metallicity will exhibit more luminous stars along the main sequence at a given
color or redder stars at a given magnitude 
(see, e.g., the isochrones of VandenBerg \& Bell 1985). An older age will go in the same
 direction for stars within a few magnitudes of the turnoff but this effect will
be much smaller. This implies that, on the upper part of a CMD, the apparent
 magnitude shift between the inner halo 
 and M4, $\Delta V$, could differ from $\Delta \mu$, the difference in
distance moduli. In general, 
 we expect:
\begin{equation}
  \Delta V = \Delta \mu + \alpha \Delta log\,t 
 + \beta \Delta {\rm [Fe/H]},
\end{equation}
 where $\alpha$ and $\beta$ are coefficients for positive 
 changes in age, $\Delta log\,t$, and metal abundance, $\Delta$[Fe/H].

 The magnitude of the correction terms
 can be obtained by appealing to isochrone calculations. 
 For this purpose, we adopt the isochrones of Montalban  et al. (2000) which
are on the HST natural system. We 
considered 11-15 Gyr isochrones for [Fe/H]$ = -1.5, -1.3,$ and $-1.0$, and
have simply taken averages of ratios of differences in $V$ at fixed $V-I$
 over differences in
$log\,t$ at a fixed [Fe/H] and over differences in [Fe/H] at a 
fixed age.

The adopted values are: 
 $\alpha \equiv dV/dlog\,t = 0.035 $,  
 $\beta \equiv  dV/d{\rm [Fe/H]} = 1.273$ so that metallicity effects
will completely dominate for modest age differences. The 
 corresponding color derivatives are 
 $d(V-I)/dlog\,t = 0.007$, and,
 $d(V-I)/d{\rm [Fe/H]} = 0.226$ so again we expect to be quite insensitive
to age effects. Because of this, in what follows below, we consider only
the more dominant metallicity effect. We expect that isochrones from other 
 sources, that
 are constrained to provide good fits to 
 observed globular cluster main sequences,
 will give similar coefficients. The results discussed below ought not to be 
 strongly dependent on the isochrone families used in this differential way.

 The main sequence associated with the inner halo
 was isolated by proper motion selection and contains the stars seen in the 
right hand
panels of Figs. 3.
 The shifts in color and magnitude between the main sequence of the 
 inner halo and M4 were determined as follows. 
 Trial values of the color shift in the range
 $ 0.01 \leq \Delta (V-I) \leq 0.09$ were selected and 
 for each value, the vertical offset was determined by
 an unweighted least-squares fit between the cluster and inner halo 
fiducials in
 the $(V,V-I)\ $ plane. The dispersion in the fit was minimized at 
 $\Delta (V-I) = 0.05$, but values of $\Delta (V-I) = 0.04$ and $ 0.06$
 were only very slightly poorer. At $\Delta (V-I) = 0.05$, the vertical
 offset derived was $\Delta V = 3.5 \pm 0.07$, 
 where the uncertainty is the
 formal 1-$\sigma$ error in the mean value. 
 The uncertainty in $\Delta V$ is mainly due to the 
 difficulty in defining the inner halo fiducial, but may also reflect a 
 slight systematic effect due to differing metallicities between the two 
 stellar populations.

Since we are not very sensitive to age effects, we 
adopt $\Delta (V-I) = 0.05$ and set $\Delta log\,t = 0$ 
 as our baseline assumption. From this we 
 obtain $\Delta {\rm [Fe/H]} = 0.22$, 
 and  $\Delta \mu =  3.2\pm 0.07$ from Eq.\,2.  With our preferred values 
 for the M4 apparent distance modulus (in the $F555W$ filter) and metallicity of
 ${\mu}_c = 12.49 \pm 0.09$ and  ${\rm [Fe/H]_c} = -1.20$ respectively
(these values are used 
 throughout this subsection), we obtain ${\mu}_b = 15.69 \pm 0.11$, and
 ${\rm [Fe/H]_b} = -0.98$ for the inner halo stars. This latter  value is in quite
good agreement with the results in Tiede \& Terndrup (1999). 
A comparison between the 
inner halo 
 main sequence, the M4 fiducial (which extends to high luminosity) and the isochrones (low luminosity only) interpolated to the derived metallicity 
 and shifted to ${\mu}_b$, is shown in Fig.\,14. 
The inner halo fit and M4 fit is quite good but of course was constructed
to be so. More satisfying is the agreement between the models and the inner halo
which indicates that our baseline assumptions are 
consistent with the data. Clearly the fit degrades towards lower luminosity where it is known that the existing isochrones do not fit intermediate metallicity stellar
sequences very well (Bedin  et al. 2002). However, we point out that in the analysis, the models
were not used in any absolute sense but only differentially to relate age and
metallicity differences to variations in color and luminosity.

Fig.\,14 demonstrates that
we have achieved reasonable concordance of the age, metallicity and distance
{\it differences} between M4 and the inner halo, through the use of the 
Montalban  et al. (2000)
isochrones. The key result from this analysis is that 
$\Delta \mu = 3.2 \pm 0.07$, which leads to $R_o/d_c = 4.37 \pm 0.13$.
As anticipated, the formal error in this ratio is  smaller than
that associated with direct estimates of either $R_o$ or $d_c$. The error
estimate does not include any uncertainty in the observed color shift, which
was fixed at $\Delta(V-I) = 0.05$ for the above estimates, 
nor does it include any allowance for
possible age differences.

\subsection{Absolute Distance Estimate to the Galactic Center}

The primary motivation for the above analysis was to obtain an
independent 
 estimate of the distance to the Galactic centre 
from the ratio of $R_o/d_c$. This, of course, 
requires a value for $d_c$.

The measurement of $R_o$, one of the classic problems of Galactic
astronomy,
 has been reviewed by Reid (1993). He 
 derived a ``best'' value of 
 $R_o = 8.0 \pm 0.5$ kpc from a weighted mean of the results from eight
 quasi-independent methods. The quoted uncertainty appears to reflect
 the range of values as much as the errors associated with each method.
More recent discussion suggests a smaller value $R_o < 7.6$ kpc (Olling
and Merrifield 2000) and the only direct measurement (Reid 1993) gives  
 $R_o = 7.2 \pm 0.7$ kpc. 

Our subdwarf based
 estimate of the distance to M4 is $d_c = 1.73 \pm 0.09$ kpc,
 which is essentially
 identical to the 
 astrometric distance (\cite{Pet95}), 
 $d_c = 1.72 \pm 0.14$ kpc.  The subdwarf distance
  is also the same as that derived from 
 an application of the Baade-Wesselink method to four cluster RR Lyrae 
 stars (\cite{LJ90}), $d_c = 1.73 \pm 0.005$ kpc (where the error is 
 due only to the dispersion in absolute magnitude among the 
 four stars used in that study).  

 The good
 agreement among the three independent estimates
 of the cluster distance together with the formally small errorbar
associated
 with the subdwarf fit motivates an estimate of $R_o$ from $d_c$. The
result is 
 $R_o = 7.6 \pm 0.2 \pm 0.4\ $ kpc,   
 where the second, systematic error now derives
 from the uncertainty in our subdwarf-based estimate of $d_c$. With the
two
 errors added absolutely, the relative error in this estimate of $R_o$
is 
 $8\%$ and, therefore, is among the most precise of the available
 estimates. If the kinematic distance to M4 is used, the error,
estimated in
 the same way, is increased to $11\%$, but the value of $R_o$ (7.6 kpc)
 may be considered a ``primary measurement'' (see Reid 1993), i.e., 
 independent of standard candles.

 A basic assumption behind the above distance estimate is that the
stellar
 density along the line of sight peaks near the tangent point, which,
with 
 $R_o = 7.5$ kpc, is $r_t=2.4$ kpc. This distance is just inside the
 nominal 2.5 kpc extent of the bar detected in the infrared
 (Dwek  et al. 1995) 
 which is associated with the {\it inner} 
 halo of the Galaxy. The bar is oriented so that the far side is 
 projected into the
 same Galactic quadrant as M4. The cluster line of sight ($b=16$\arcdeg) 
 is fairly high, so
 the bar, as modeled by \cite{Dw95}, falls below our 
 tangent point and should not be a
 factor in this analysis. Nevertheless, poor knowledge of the
 spatial distribution 
 of the stellar populations in the inner parts of the Galaxy does
 introduce additional systematic uncertainty in our results. An example
of this
may be a red population obvious in the CMD of the inner halo. This
population
can be seen in the right-hand panels of Fig. 3 as a redward extension of
the main
sequence beginning at about $V = 23$. The kinematics of this population
indicates
that its proper motion is consistent with that of the inner halo (i.e.,
distinct from that of M4)
 but this is not a very strong statement as the small
motion of this sample with respect to an extragalactic rest frame is
also
similar to what one would expect from nearby disk or thick disk stars
(see 
Kalirai  et al. 2003 for further discussion).

 The methodology described here is applicable to the inner halo observed 
 behind any foreground cluster and is capable of considerable refinement
with
 larger photometric data sets and  better stellar and Galactic structure 
 models. Given the present context, we simply note that the observed 
 inner halo sequence is fully consistent
 with our adopted cluster distance and the generally accepted values for
$R_o$.

\end{appendix}

\begin{acknowledgements}
The research of HBR is supported in part by the Natural Sciences
and Engineering Research Council of Canada. HBR extends his appreciation to the Killam Foundation and
the Canada Council for the award of a Canada Council Killam Fellowship. RMR and MS acknowledge support from proposal GO-8679 and BH from a Hubble Fellowship
HF-01120.01 both of which were provided by NASA through a grant from the Space Telescope Science Institute which is operated by AURA under NASA contract NAS5-26555. BKG acknowledges the support of the Australian Research Council through its Large Research Grant Program A00105171.
\end{acknowledgements}

\clearpage

\clearpage

\figcaption{The variation in the sky background as a function of time (Mean Julian Days) or angle between the Sun and M4 on the long exposure frames
in the outer annulus
in both $F606W$ and $F555W$. The decrease is due to the lowering of the zodiacal light background as the Sun-M4 angle increased. The few jumps in the brightness levels are due to observations taken close to the bright limb of the Earth.}\label{fig1}

\figcaption{Total proper motion difference between M4 and background/foreground objects as a function of magnitude for the outer field data. In this diagram the motion is zero-pointed on M4 and the displacement is given in HST pixels over the 6 year baseline of the observations. In order to separate the cluster from the field all objects possessing motion within 0.5 pixel of that of M4 were assumed to be cluster members. Because we do not resolve the M4 internal motion, it is clear from the diagram that our uncertainties in the measured proper motions do not exceed 2 mas/yr down to $F606W = 27.5$ and at $F606W = 29.5$ they are smaller than 7 mas/yr.}\label{fig2}

\figcaption{All stars within the 4 annuli selected by proper motion
are plotted in the sequence $a-d$ with $a$ being the inner annulus.
The upper section of each plot is the proper motion selection with the
units being relative motion in HST pixels (0.1") over the 6 year time baseline.
The densest and most concentrated clump consists of the M4 stars and 
their CMD is
illustrated in the left hand panels. The right hand sequence is largely due
 to stars in the inner halo of the Galaxy.
 The cluster main sequence
 is the reddest of the sequences observed in these plots, the cluster 
 white dwarfs are the bluest sequence.}\label{fig3}
  
\figcaption{The magnitude of the average photometric errors in the inner annuli
(1-3) in both $I$ and $V$ from artificial star tests are plotted as a function of stellar magnitude.}\label{fig4}

\figcaption{CMDs in the 4 annuli with the cluster main sequence fiducial
shifted brighter by 0.75 mags in order to indicate the location of an equal-mass binary sequence. Saturation is important for $F555W < 20$ particularly in the outer field. It appears from this diagram that the equal mass cluster binary frequency is very low.}\label{fig5}

\figcaption{A simulation of the binary frequency in our fields. The stars were drawn at random from mass functions with the indicated slopes and assigned errors as in Fig. 4. Figures 5 and 6 collectively suggest a modest main sequence binary frequency in our M4 fields.} \label{fig6}

\figcaption{Histograms of the deviation of M4 stars (in magnitudes) from the main-sequence fiducial. Small
excesses are clearly seen towards negative residuals at about 0.75 magnitudes
in the inner 2 annuli suggesting a photometric main sequence
binary frequency of a few percent.}\label{fig7}

\figcaption{Observed cluster luminosity functions (points with error bars) in $V$ compared with the theoretical function derived from the best fitting mass function. The mass function slopes ($\alpha_{Salpeter} = +2.3$) and $\chi^2$ fit between theory and observed function are indicated.}\label{fig8}

\figcaption{As in Fig. 8 except that the functions are developed from the $I$ magnitudes and are independent of those in $V$. The agreement in mass function slopes between the two bandpasses in the same annuli is remarkably good.}\label{fig9}

\figcaption{The M4 surface brightness profile defined
 by counts of stars with $V < 16$ 
 (symbols with error bars) is compared to the
 profile of a corresponding mass-class from the MMK model adopted here. 
 The best fit occurs for a scale radius, $r_s = 55{\arcsec}$.}\label{fig10}

\figcaption{The  mass functions predicted by the MMK model in our 4 annuli 
are shown. The mass function in annulus 4 (A4) reproduces the input
global mass function, except for the very highest masses, which were 
not observed here and are predicted to be 
somewhat segregated toward the cluster center. 
The calculated mass functions in
the inner annuli are not power laws but, nevertheless,
are crudely consistent with 
the power law slopes derived in \S 5 of the text. A comparison of the model 
predictions to the observed luminosity functions is 
provided in Fig. 12. }\label{fig11}

\figcaption{Model luminosity functions in $V$ in the 4 annuli compared to the data (points with error bars). This diagram indicates that we have achieved reasonable agreement between the M4 luminosity functions and those expected in equilibrium MMK models.}\label{fig12}

\figcaption{As in Fig. 12 except that the luminosity functions here are
derived from the $I$-band data.}\label{fig13}

\figcaption{Left hand panel: The M4 CMD in the inner 3 annuli. Right hand panel: The CMD of non-cluster stars (largely inner halo) in the direction of M4 from the 3 inner annuli. This inner halo main sequence is compared to the M4 fiducial (the fiducial has been extended to brighter magnitudes using ground-based data) and the isochrones (bluest at low luminosity) interpolated to the derived metallicity 
and both shifted to a distance modulus of 3.2 magnitudes larger than that of the cluster.}\label{fig14}

\clearpage

\begin{deluxetable}{lllr}
\tablewidth{0pc}
\tablecaption{M4 Physical Parameters \label{table1}}
\tablehead{
\colhead{Parameter} & \colhead{Value} & \colhead{Comment} 
 & \colhead{Reference} 
 }
\startdata

distance (kpc) & 1.73 $\pm$ 0.09 & Subdwarf fit & 1\\
\phm{distance} & 1.72 $\pm$ 0.14 & Astrometric & 2 \\
\phm{distance} & 1.72 $\pm$ 0.01 & Baade-Wesselink & 3 \\
true distance modulus & 11.18 & as above & 1,2,3 \\
\\
%\tablevspace{12pt}
E(V-I)  & 0.51 $\pm$ 0.02       & Subdwarf fit (Johnson Colors)  & 1 \\
R$_{\rm V}$     & 3.8 & Observed extinction (Johnson Colors)  & 2,3,4,5 \\
A$_{\rm V}$     & 1.33  & Model extinction curves (Johnson Colors) & 6 \\
A$_{\rm F555W}$     & 1.31  & Model extinction curves & 6 \\
A$_{\rm F814W}$     & 0.82  & Model extinction curves & 6 \\

[Fe/H]          & $-1.20$ $\pm$ 0.15 & Spectroscopy & 7 \\
\\
%\tablevspace{12pt}
core radius ($r_c$)     & 53$\arcsec$   & King model  & 8,11 \\
concentration ($c$) & 1.74      & King model  & 10 \\
half light radius ($l_{rh}$)     & 269$\arcsec$   & King model  & 8 \\
half mass radius ($m_{rh}$)     & 367$\arcsec$   & King model  & 11 \\
integrated luminosity (${\cal L_V}$) & 6.25$\times 10^4\ L_{\odot}$
    & photometry & 9,11 \\
central Mass/Light $({\cal M}/{\cal L})_0$ & 1.64 & King model & 11 \\
global Mass/Light  $({\cal M}/{\cal L})$ & 1.00 & King Model & 10, 11 \\
relaxation time ($t_{rh}$) & 2.18$\times 10^8$ yr. 
    & Spitzer formula & 9,11 \\
central relax. time ($t_{rc}$)
    & 1.80$\times 10^7$ yr. & Spitzer formula & 9,11 \\
\enddata
\tablerefs{(1) Richer {\it et al.} 1997 (Paper I); (2) Peterson {\it et al.} 1995; 
(3) Liu \& Janes 1990; (4) Dixon \& Longmore 1993); (5) Vrba {\it et al.} 1993;
 (6) Cardelli {\it et al.} 1989; (7) Drake {\it et al.} 1994; (8) Trager {\it et al.} 1993; 
(9) Djorkovski 1993; (10) Pryor \& Meylan 1993;
(11) this paper.}
\end{deluxetable}

\clearpage

\begin{deluxetable}{ccccc}
\tablecolumns{5}
\tablewidth{0pc}
\tablecaption{Field Geometry \label{table2}}
\tablehead{
\colhead{Annulus} &  \colhead{inner} & \colhead{outer} &
 \colhead{median} & \colhead{area} \\
\colhead{} & \colhead{$r_c$} & 
\colhead{$r_c$} & 
 \colhead{\arcmin} & \colhead{\sq \arcmin}  
}

\startdata
1 & 0 &1.5 & 0.938 & 1.75 \\
2 & 1.5 & 2.5 & 1.625 & 2.87 \\
3 & 2.5 & 4.0 & 2.522 & 2.73 \\
4 & 4.0 & \nodata & 5.028 & 4.98 \\
\enddata
\end{deluxetable}

\clearpage

\begin{deluxetable}{rrrrrrl}
\footnotesize
\tablecaption{Exposure Record}
\tablewidth{0pt}
\tablehead{
\colhead{Field($r_c$)} & \colhead{Date} & \colhead{Program ID} & \colhead{Filter} &
\colhead{Exposure} & \colhead{Number of frames}  & \colhead{Comments}
}
\startdata
1 & 1995 & 5461   & F336W & 1300 & 1   &  \\	
1 & 1995 & 5461   & F336W & 1500 & 7   &  \\	
1 & 1995 & 5461   & F555W & 1000 & 15  &  \\	
1 & 1995 & 5461   & F814W & 600  & 1   & first epoch for pm  \\	
1 & 1995 & 5461   & F814W & 700  & 7   & first epoch for pm \\	
1 & 2000 & 8153   & F814W & 700  & 8   & second epoch for pm \\
  &      &        &      &      &     &  \\
2 & 1995 & 5461   & F336W & 1300 & 1   & \\	
2 & 1995 & 5461   & F336W & 1500 & 7   &  \\	
2 & 1995 & 5461   & F555W & 1000 & 15  &  \\	
2 & 1995 & 5461   & F814W & 600  & 1   & first epoch for pm  \\	
2 & 1995 & 5461   & F814W & 700  & 7   & first epoch for pm  \\	
2 & 2000 & 8153   & F814W & 700  & 8   & second epoch for pm \\
  &      &        &      &      &     &  \\
6 & 1995 & 5461   & F555W & 2100 & 15  & first epoch for pm \\	
6 & 1995 & 5461   & F814W & 800  & 9   & first epoch for pm \\	
6 & 2001 & 8679   & F606W & 1300 & 98  & second epoch for pm \\	
6 & 2001 & 8679   & F814W & 1300  & 148 & second epoch for pm \\	

\enddata

\end{deluxetable}

\clearpage
\begin{deluxetable}{cccccc}
\tablecolumns{6}
\tablewidth{0pc}
\tablecaption{Annulus 1: Photometry and Astrometry\tablenotemark{a}} %\label{table4}}
\tablehead{
\colhead{RA} &  \colhead{DEC} & \colhead{F555W} &
 \colhead{F814W} & \colhead{$\mu_{\alpha}cos(\delta)$} & \colhead{$\mu_{\delta}$} \\
}

\startdata
245.8968 & -26.5372 & 19.712 & 18.236 & -13.740 & -19.528 \\ 
245.8971 & -26.5371 & 19.615 & 18.139 & -14.558 & -20.437 \\ 
245.8977 & -26.5388 & 23.041 & 20.449 & -12.390 & -18.678 \\ 
245.8978 & -26.5365 & 21.049 & 19.136 & -12.431 & -17.723 \\ 
245.8978 & -26.5386 & 22.851 & 20.470 & -14.528 & -20.200 \\ 
\enddata
\tablenotetext{a}{This is just the first 5 entries in this table which can
be retrieved in full from the Astronomical Journal Archives.}
\end{deluxetable}

\clearpage
\begin{deluxetable}{cccccc}
\tablecolumns{6}
\tablewidth{0pc}
\tablecaption{Annulus 2: Photometry and Astrometry\tablenotemark{a}} %\label{table5}}
\tablehead{
\colhead{RA} &  \colhead{DEC} & \colhead{F555W} &
 \colhead{F814W} & \colhead{$\mu_{\alpha}cos(\delta)$} & \colhead{$\mu_{\delta}$} \\
}

\startdata
245.9035 & -26.5467 & 21.001 & 19.157 & -11.547 & -17.889 \\ 
245.9036 & -26.5457 & 23.862 & 21.316 & -14.084 & -18.983 \\ 
245.9041 & -26.5464 & 20.320 & 18.630 & -11.496 & -19.782 \\ 
245.9042 & -26.5465 & 23.886 & 22.975 & -11.920 & -19.095 \\ 
245.9050 & -26.5456 & 20.600 & 18.817 & -11.154 & -19.344 \\ 
\enddata
\tablenotetext{a}{This is just the first 5 entries in this table which can
be retrieved in full from the Astronomical Journal Archives.}
\end{deluxetable}

\clearpage
\begin{deluxetable}{cccccc}
\tablecolumns{6}
\tablewidth{0pc}
\tablecaption{Annulus 3: Photometry and Astrometry\tablenotemark{a}} %\label{table6}}
\tablehead{
\colhead{RA} &  \colhead{DEC} & \colhead{F555W} &
 \colhead{F814W} & \colhead{$\mu_{\alpha}cos(\delta)$} & \colhead{$\mu_{\delta}$} \\
}

\startdata
245.9183 & -26.4944 & 24.773 & 22.198 & -12.803 & -17.138 \\ 
245.9184 & -26.4946 & 23.042 & 20.727 & -13.528 & -16.526 \\ 
245.9194 & -26.4951 & 20.653 & 18.888 & -13.127 & -17.936 \\ 
245.9196 & -26.4956 & 22.404 & 20.254 & -11.859 & -16.632 \\ 
245.9200 & -26.4943 & 27.027 & 25.723 & -13.396 & -17.554 \\ 
\enddata
\tablenotetext{a}{This is just the first 5 entries in this table which can
be retrieved in full from the Astronomical Journal Archives.}
\end{deluxetable}

\clearpage
\begin{deluxetable}{cccccc}
\tablecolumns{6}
\tablewidth{0pc}
\tablecaption{Annulus 4: Photometry and Astrometry\tablenotemark{a}} %\label{table7}}
\tablehead{
\colhead{RA} &  \colhead{DEC} & \colhead{F555W} &
 \colhead{F814W} & \colhead{$\mu_{\alpha}cos(\delta)$} & \colhead{$\mu_{\delta}$} \\
}

\startdata
245.9678 & -26.5533 & 22.027 & 20.140 & -11.631 & -19.955 \\ 
245.9680 & -26.5538 & 22.111 & 20.044 & -11.374 & -19.479 \\ 
245.9695 & -26.5511 & 25.209 & 24.284 & -13.151 & -16.950 \\ 
245.9695 & -26.5516 & 28.134 & 26.669 & -8.621  & -14.968 \\ 
245.9701 & -26.5515 & 22.273 & 20.152 & -13.651 & -19.347 \\ 
\enddata
\tablenotetext{a}{This is just the first 5 entries in this table which can
be retrieved in full from the Astronomical Journal Archives.}
\end{deluxetable}
\clearpage

\begin{deluxetable}{cr}
\footnotesize
\tablecolumns{2}
\tablewidth{0pc}
\tablecaption{Main Sequence Fiducial\label{table8}}
\tablehead{
\colhead{$F555W - F814W$} & \colhead{$F555W$}}

\startdata
     1.384  &  19.368\\
     1.455  &  19.650\\
     1.525  &  19.888\\
     1.596  &  20.102\\
     1.667  &  20.312\\
     1.737  &  20.525\\
     1.808  &  20.749\\
     1.878  &  20.992\\
     1.949  &  21.264\\
     2.020  &  21.563\\
     2.090  &  21.885\\
     2.161  &  22.224\\
     2.232  &  22.567\\
     2.302  &  22.901\\
     2.373  &  23.228\\
     2.444  &  23.542\\
     2.514  &  23.831\\
     2.585  &  24.086\\
     2.655  &  24.320\\
     2.726  &  24.550\\
     2.797  &  24.783\\
     2.867  &  25.020\\
     2.938  &  25.268\\
     3.009  &  25.532\\
     3.079  &  25.806\\
     3.150  &  26.048\\
     3.221  &  26.238\\
     3.291  &  26.394\\
     3.362  &  26.538\\
     3.432  &  26.687\\
     3.503  &  26.843\\
\enddata
\end{deluxetable}

\clearpage

\begin{deluxetable}{cccc}
\tablecolumns{4}
\tablewidth{0pc}
\tablecaption{Upper Limits on M4 Binary Fractions \label{table9}}
\tablehead{
\colhead{Annulus} & \colhead{$f_b$ (\%)} & \colhead{$\pm \sigma$} &
        \colhead{Number\tablenotemark{a}}  
 }
\startdata
1 & 2.2 & 0.8 & 403 \\
2 & 1.1 & 0.4 & 805 \\
3 & 1.1 & 0.3 & 759 \\
4 & 1.8 & 0.5 & 514 \\
\enddata
\tablenotetext{a}{The total number of main sequence stars used.}
\end{deluxetable}

\clearpage

\begin{deluxetable}{ccc}
\tablecolumns{3}
\tablewidth{0pc}
\tablecaption{Mass Function Slopes at Different Radii}
\tablehead{
\colhead{Annulus} & \colhead{$r/r_h$\tablenotemark{a}} & \colhead{$\alpha_{observed}$} \\
}
\startdata
1 & $0.17 \pm 0.09$ & $ -0.7$ \\
2 & $0.30 \pm 0.09$ & $ -0.4 $ \\
3 & $0.47 \pm 0.14$ & $ -0.2$ \\
4 & $0.94 \pm 0.19$ & $  0.1 $ \\
\enddata
\tablenotetext{a}{$r_h = 269\arcsec .2$ from Trager  et al. (1993)}
\label{mftab}
\end{deluxetable}

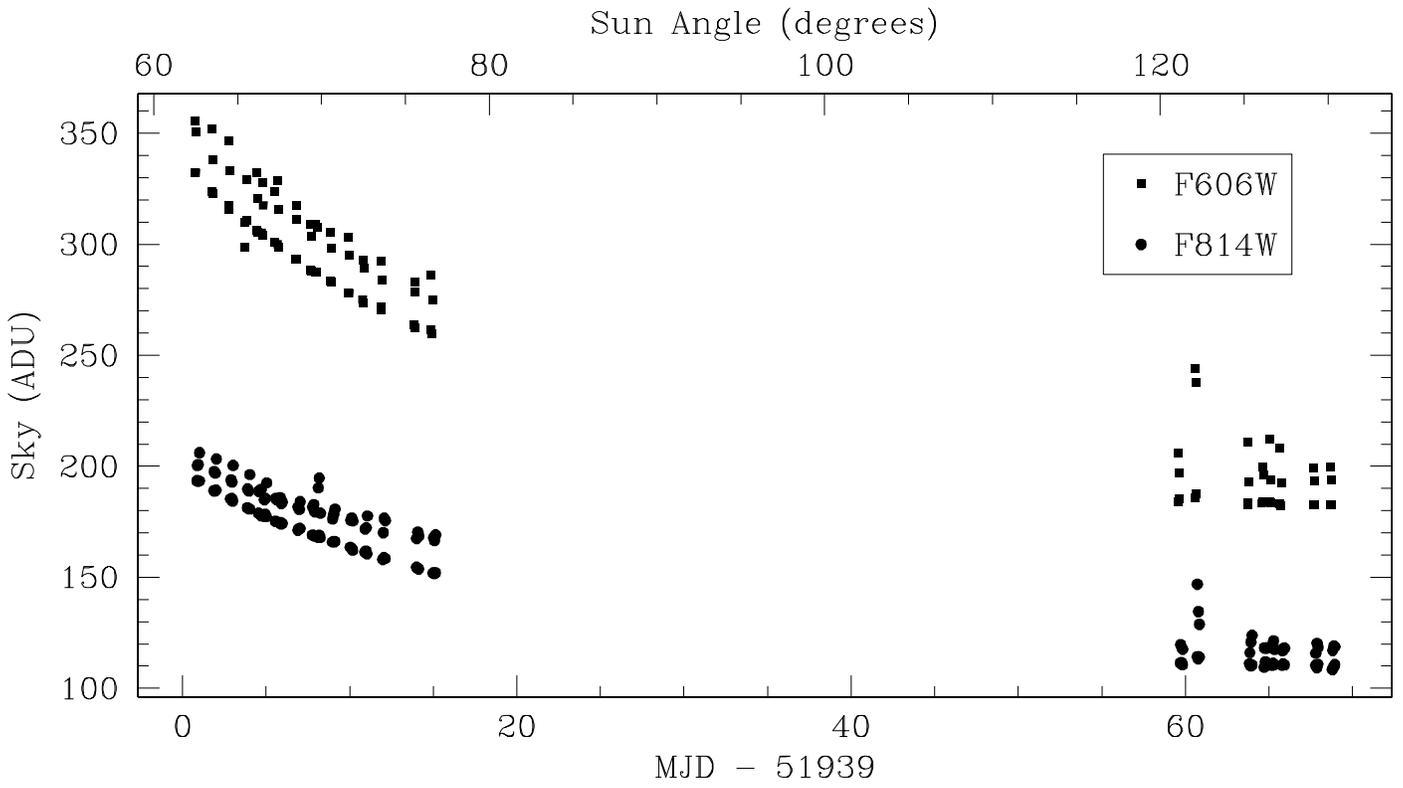
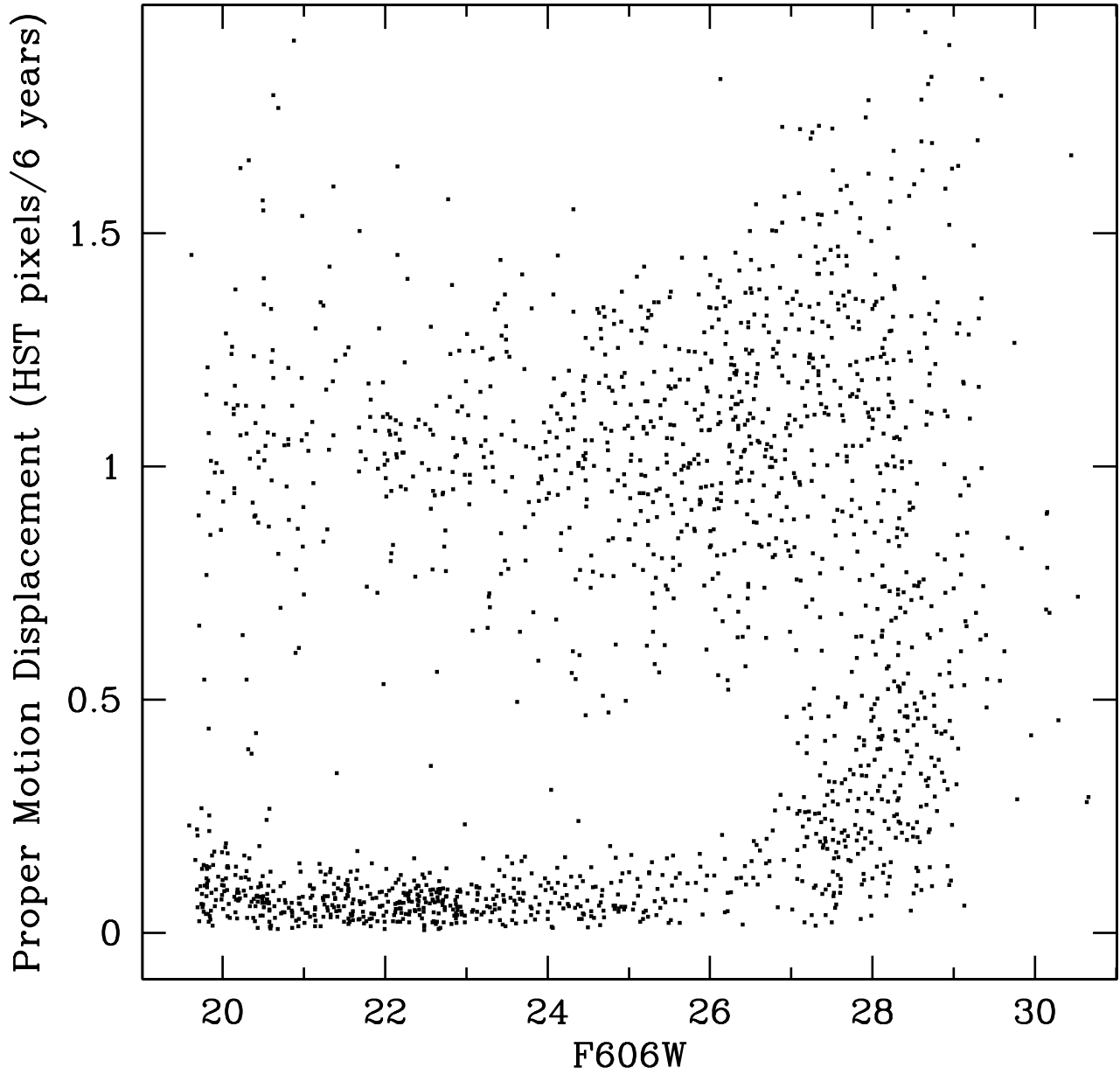
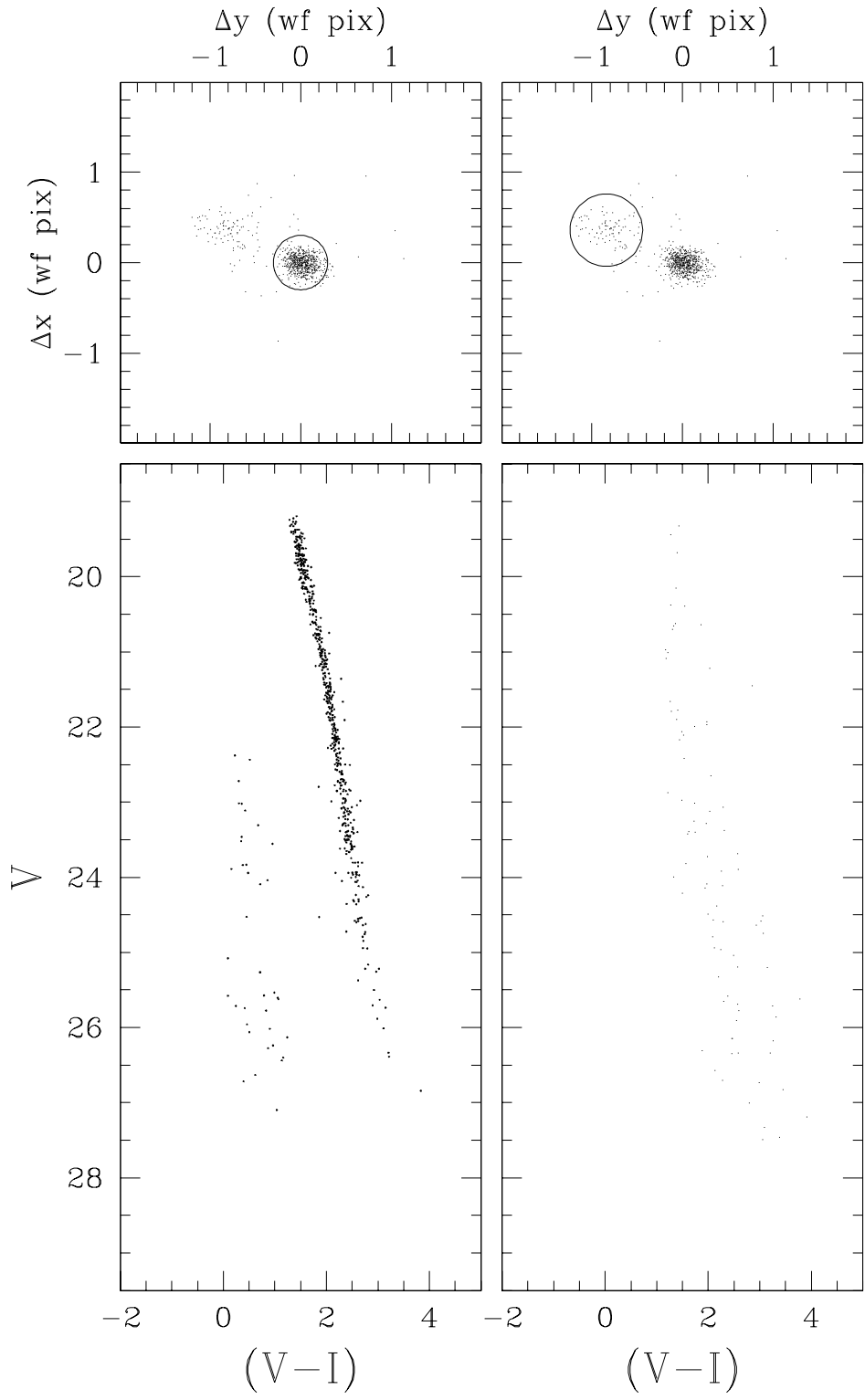
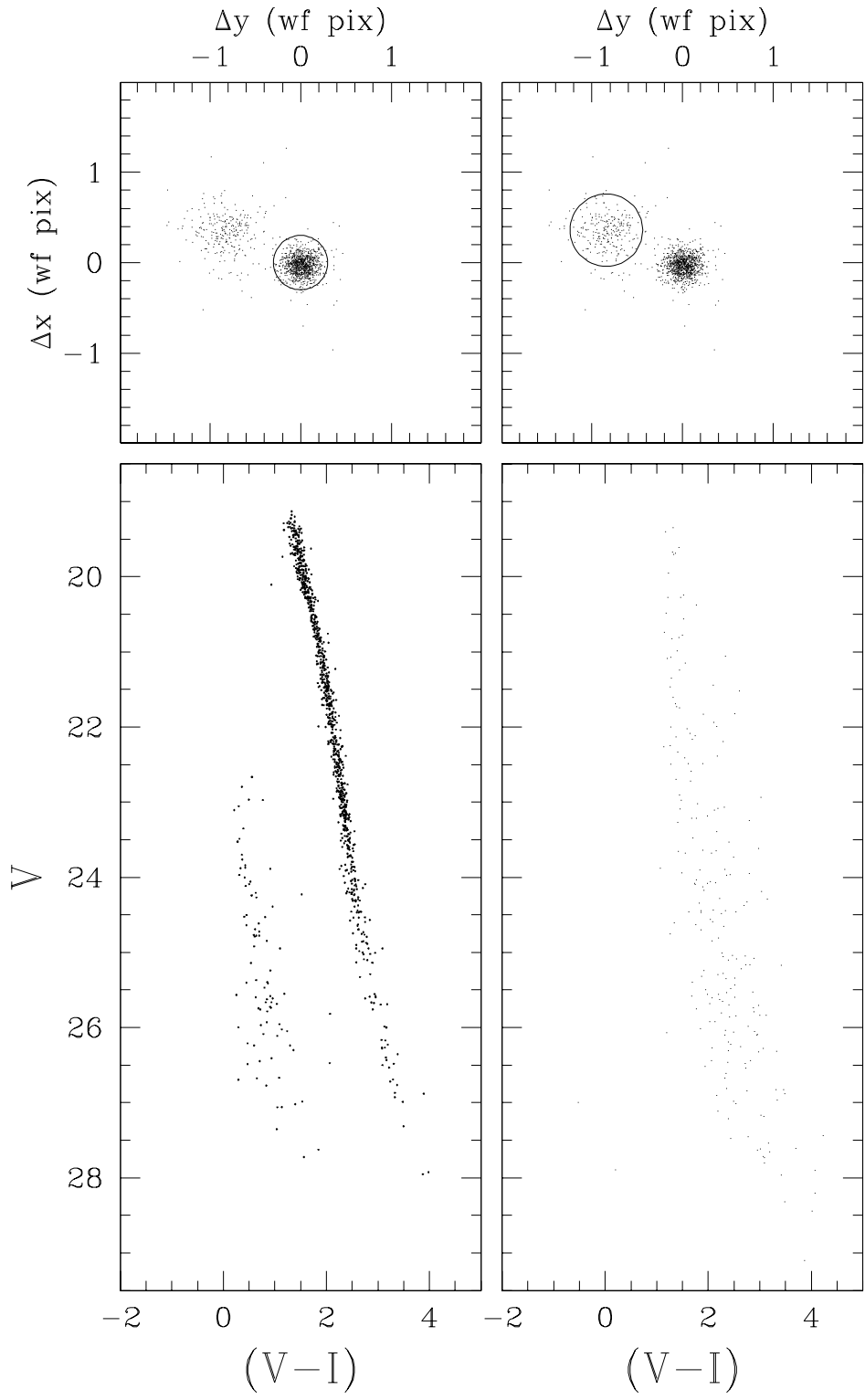
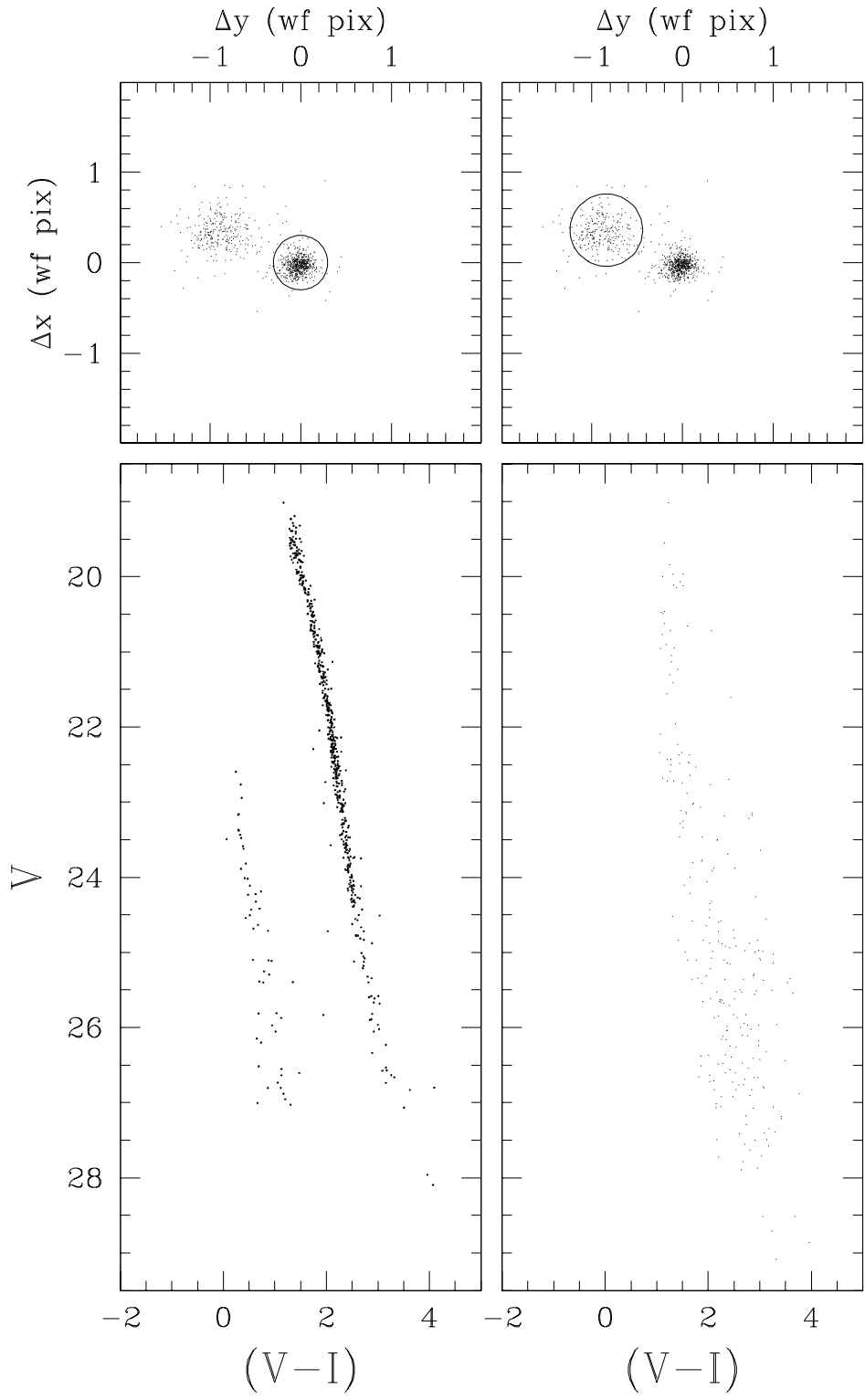
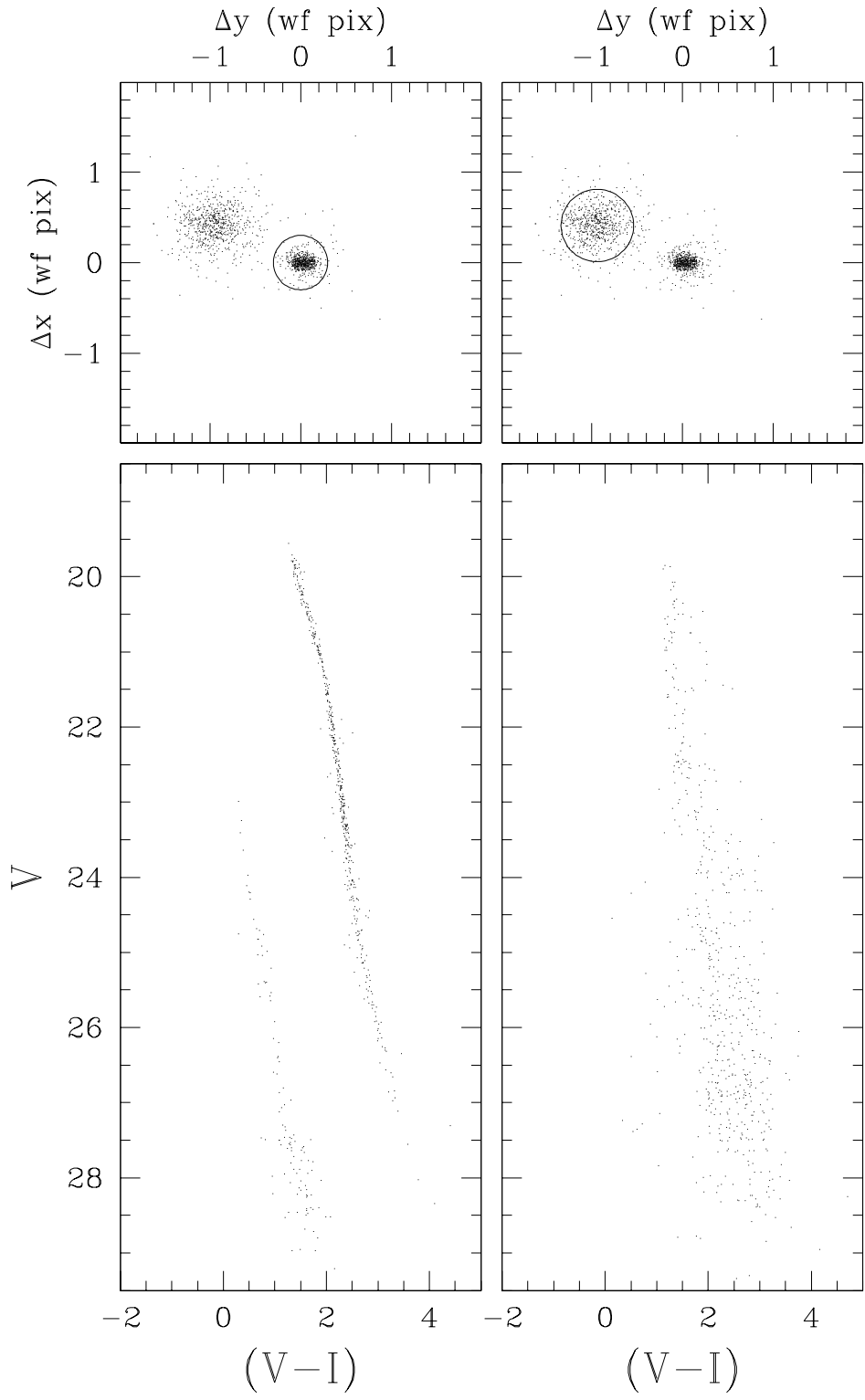
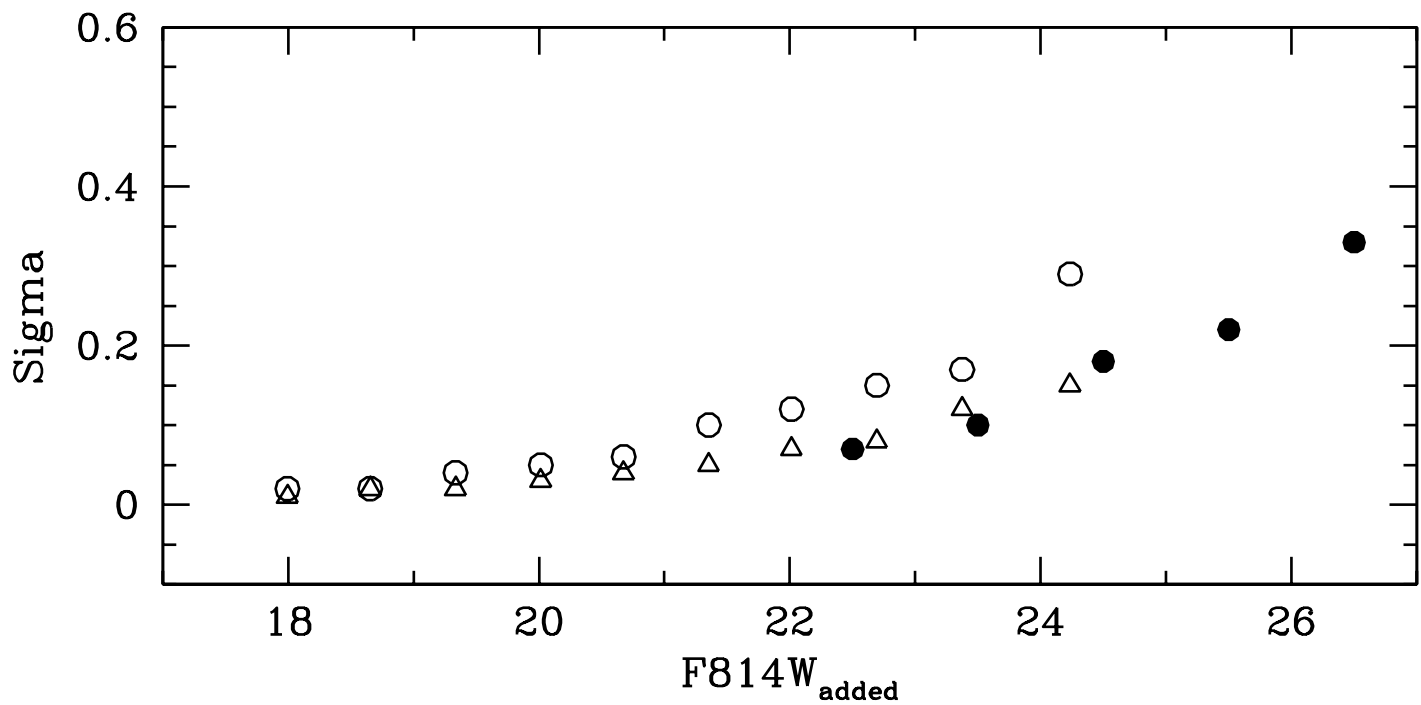
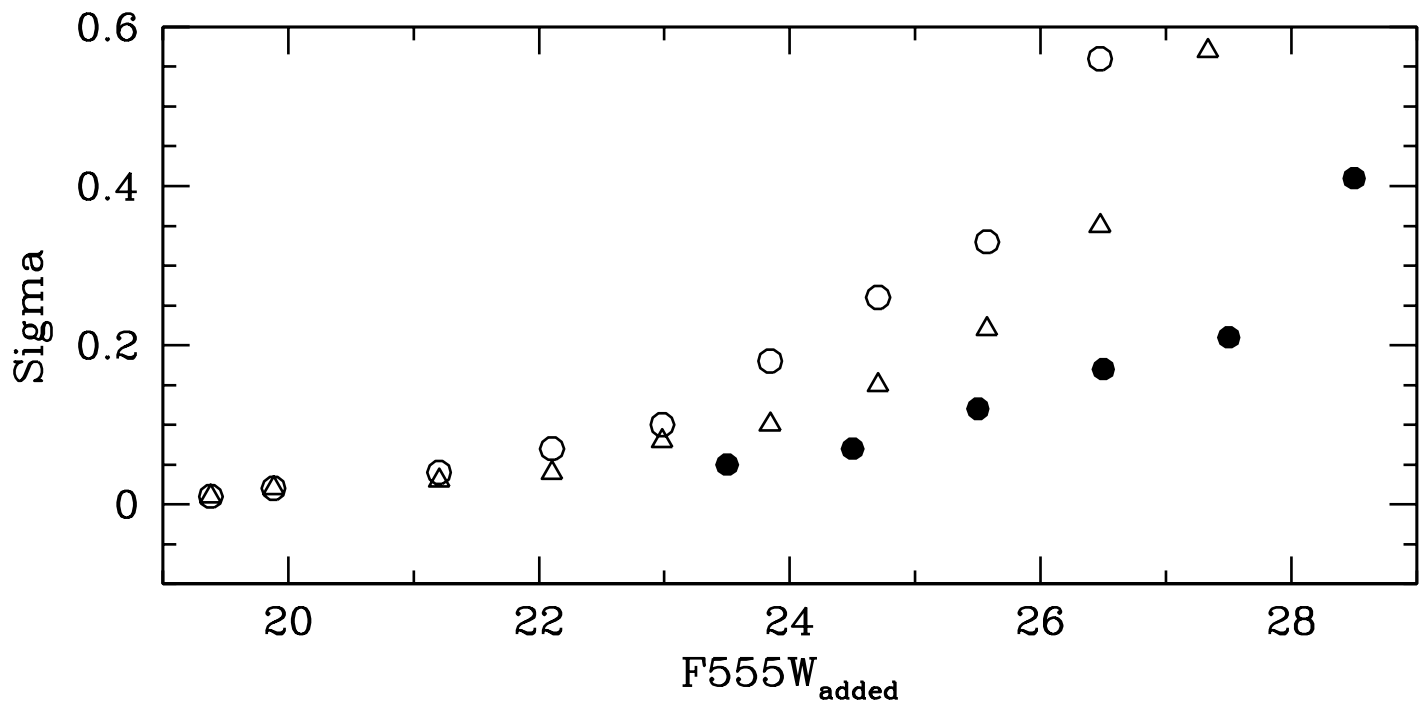
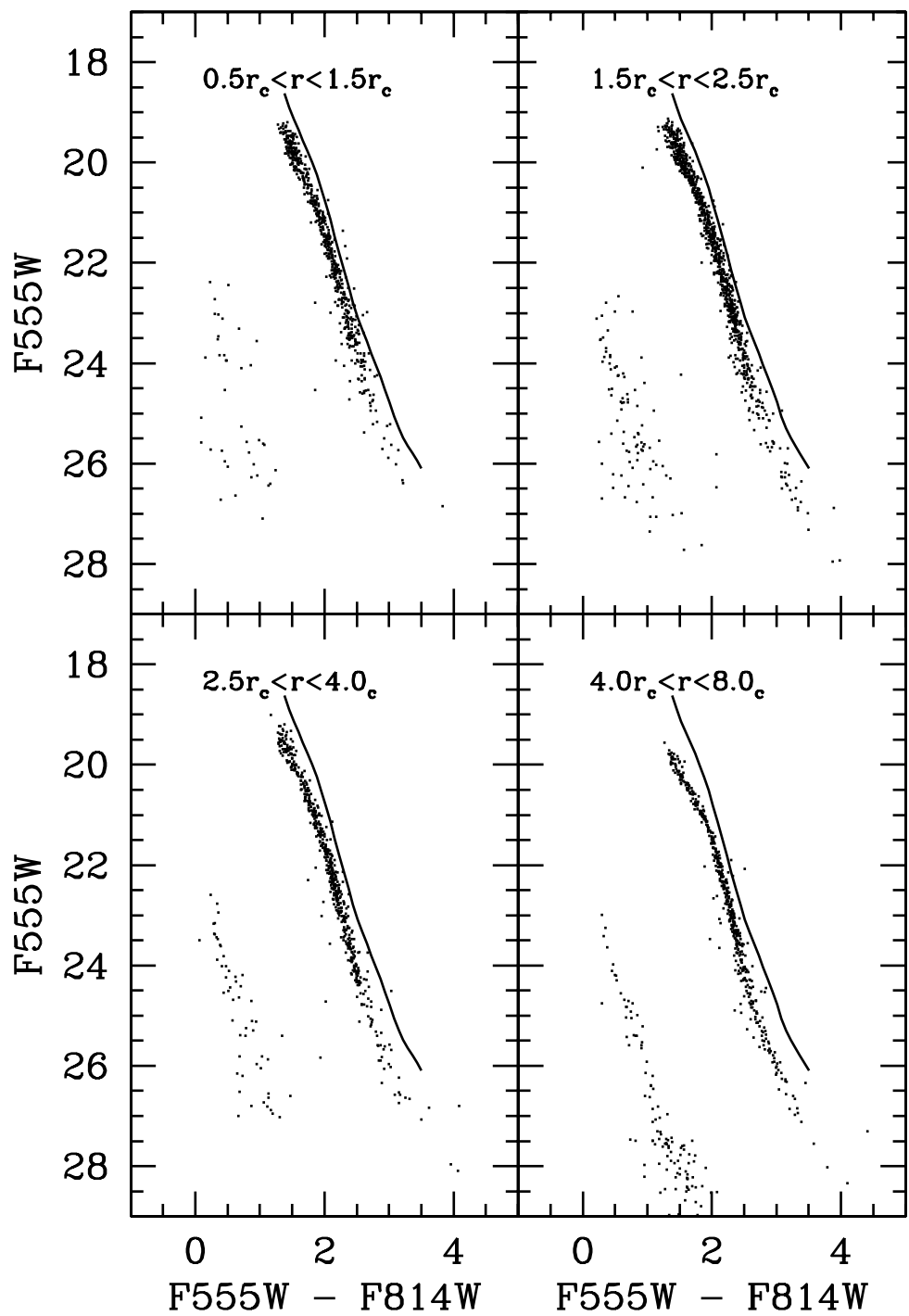
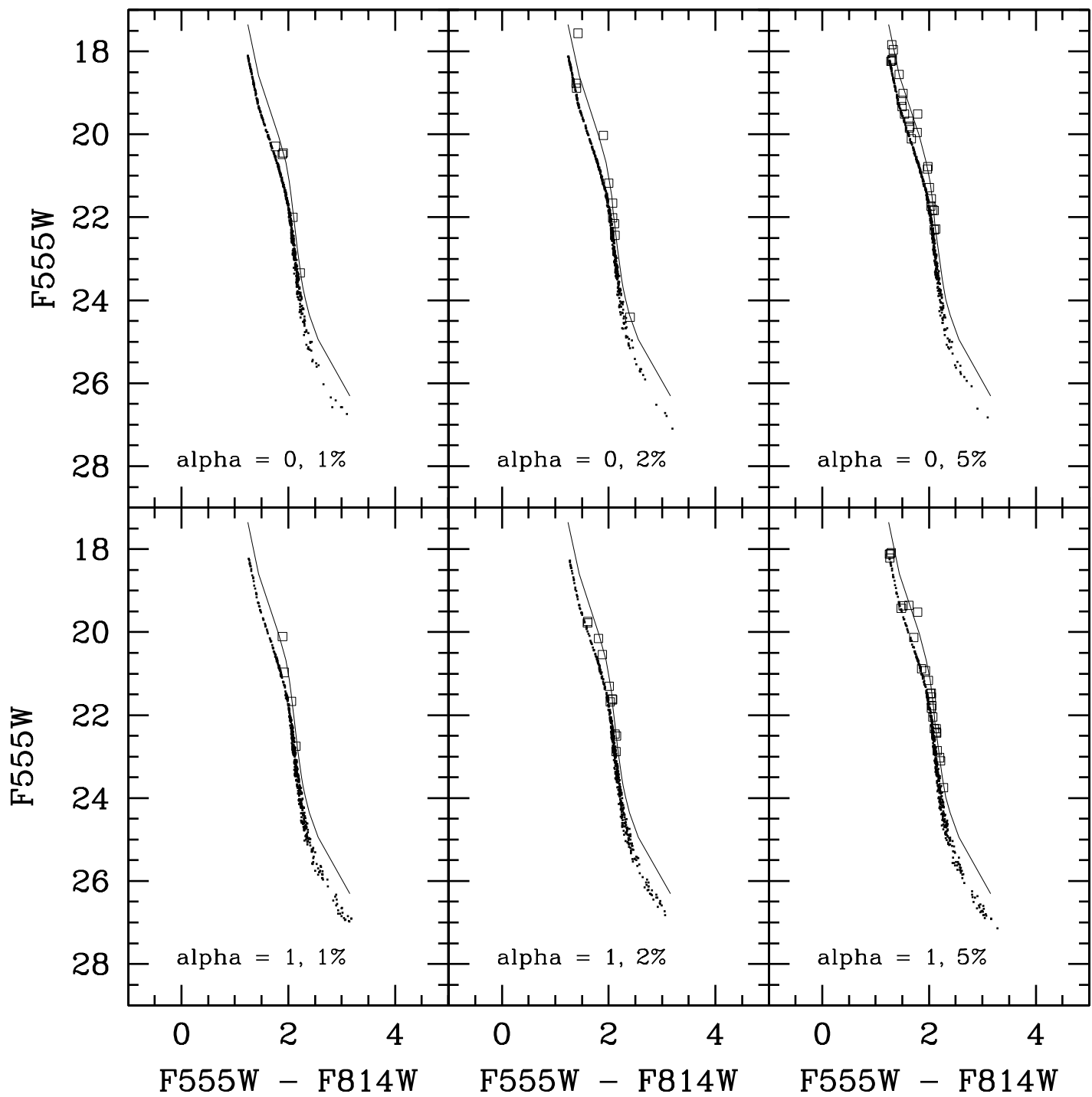
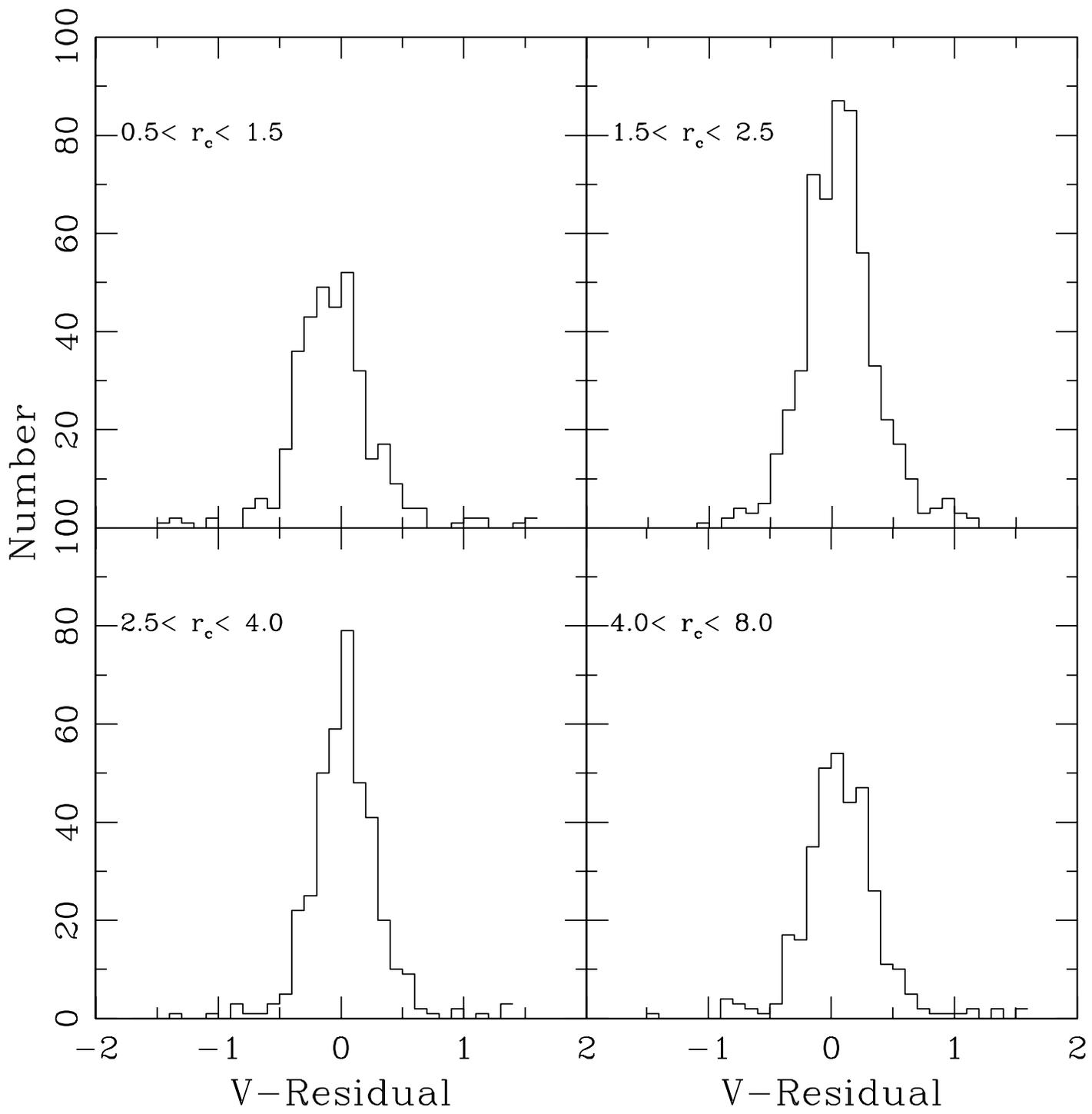
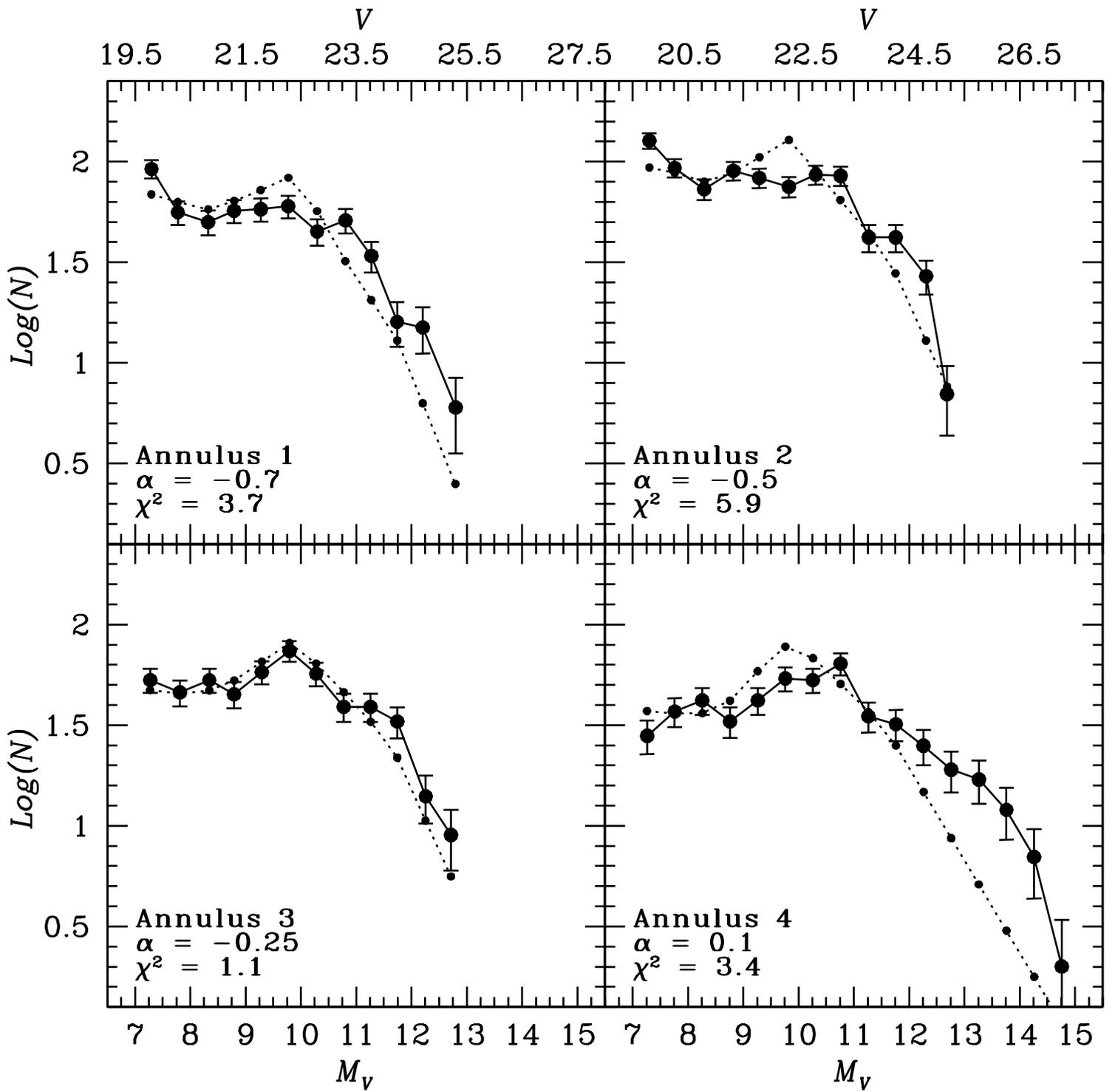
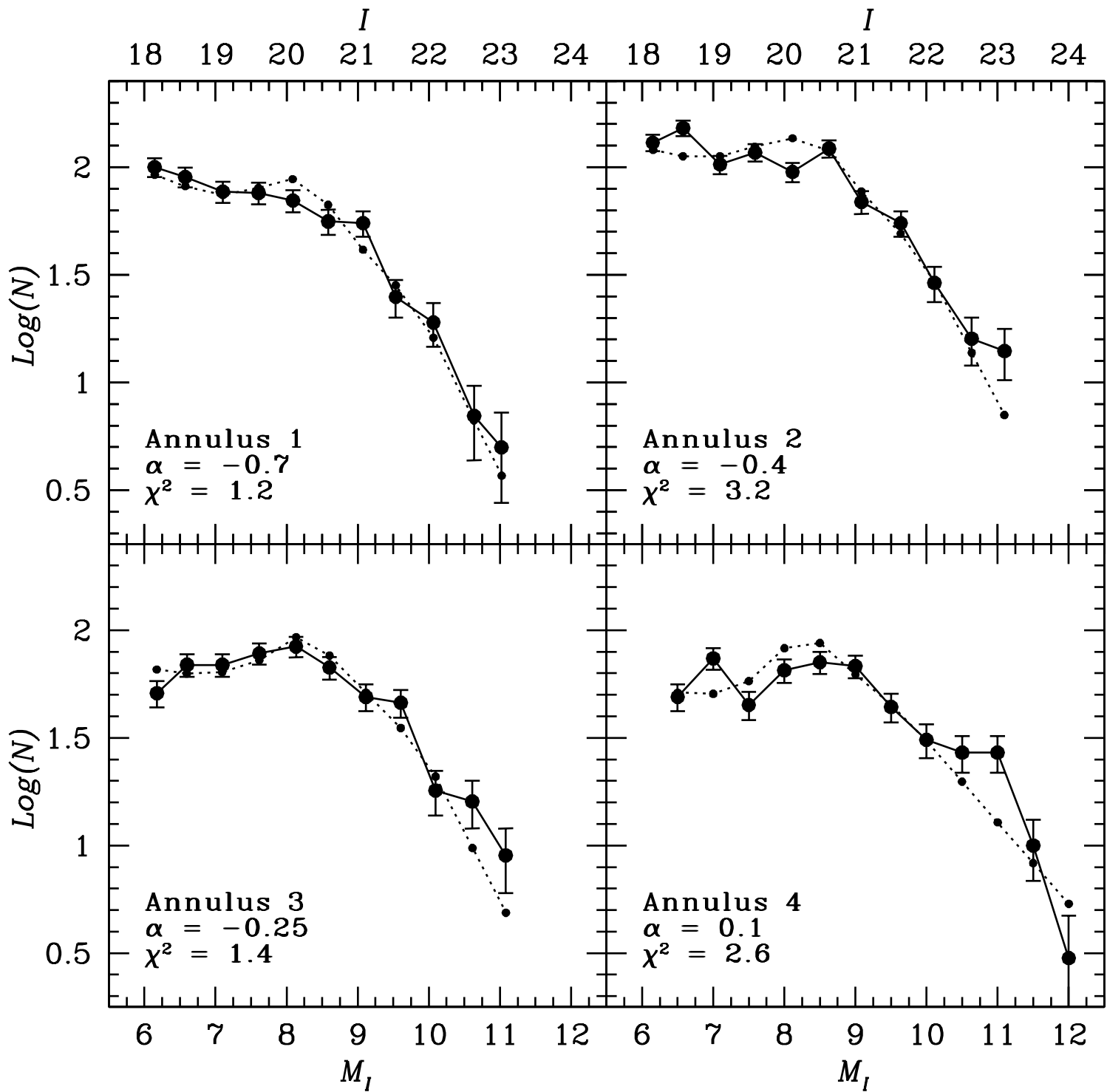
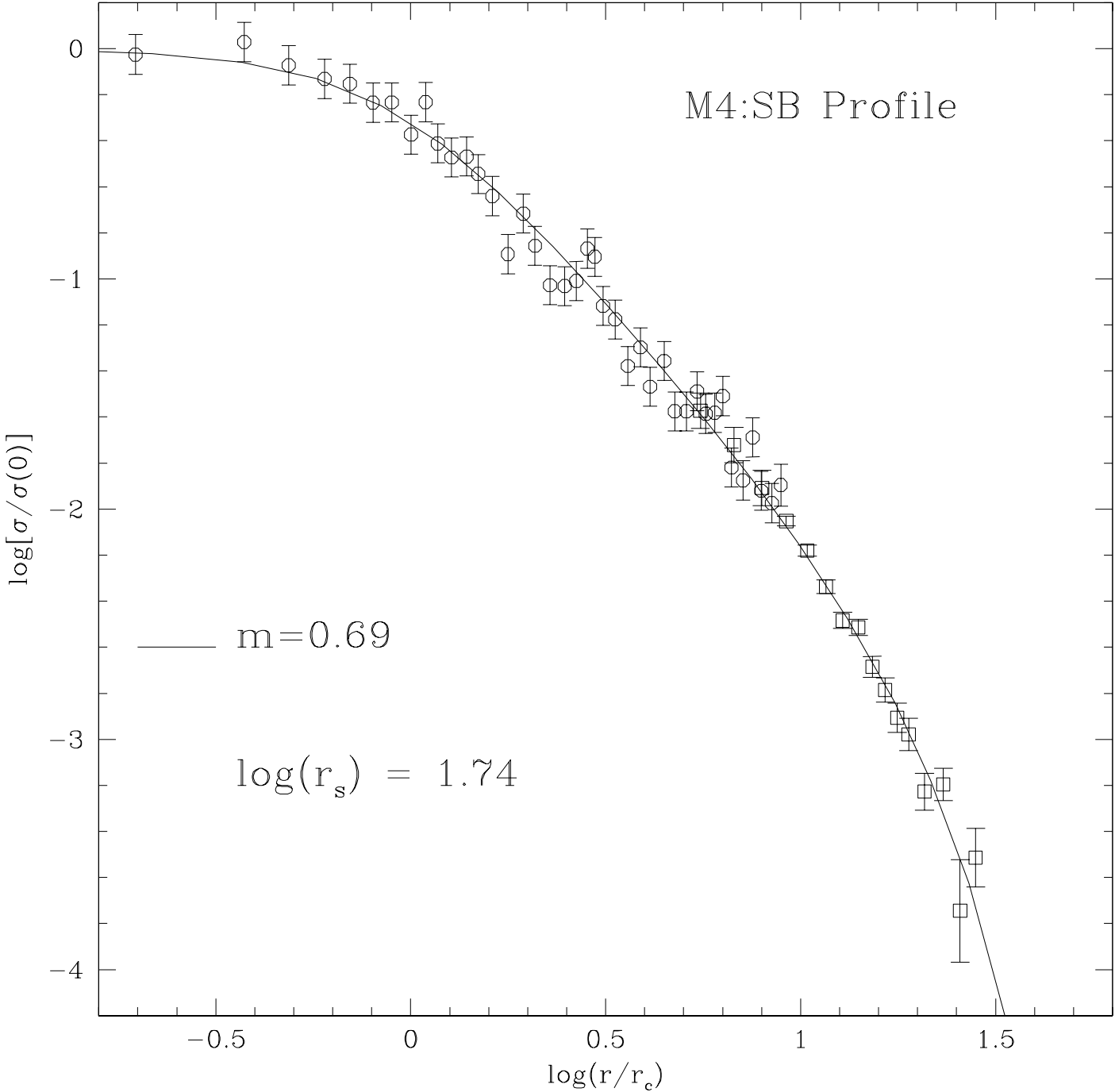
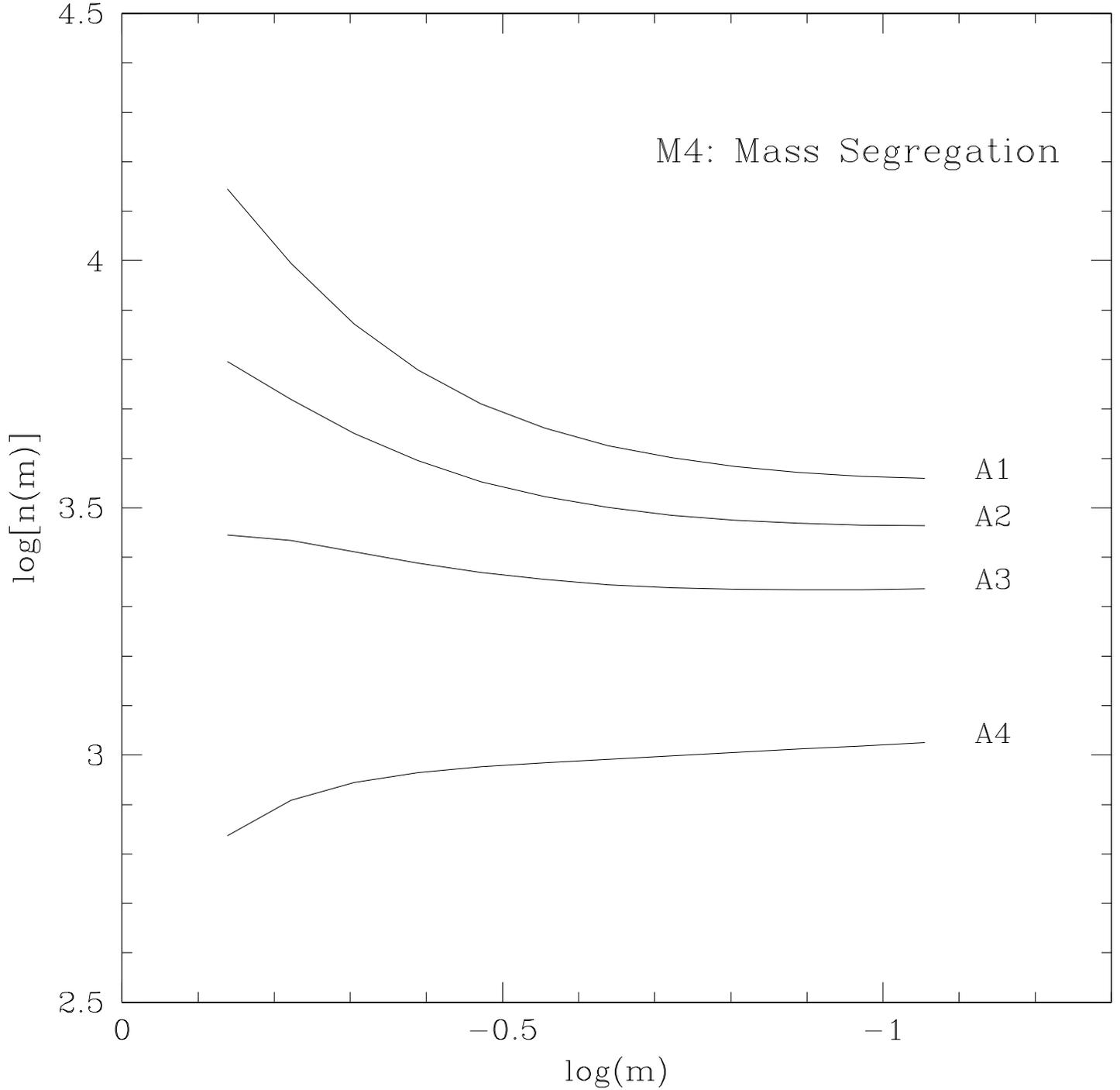
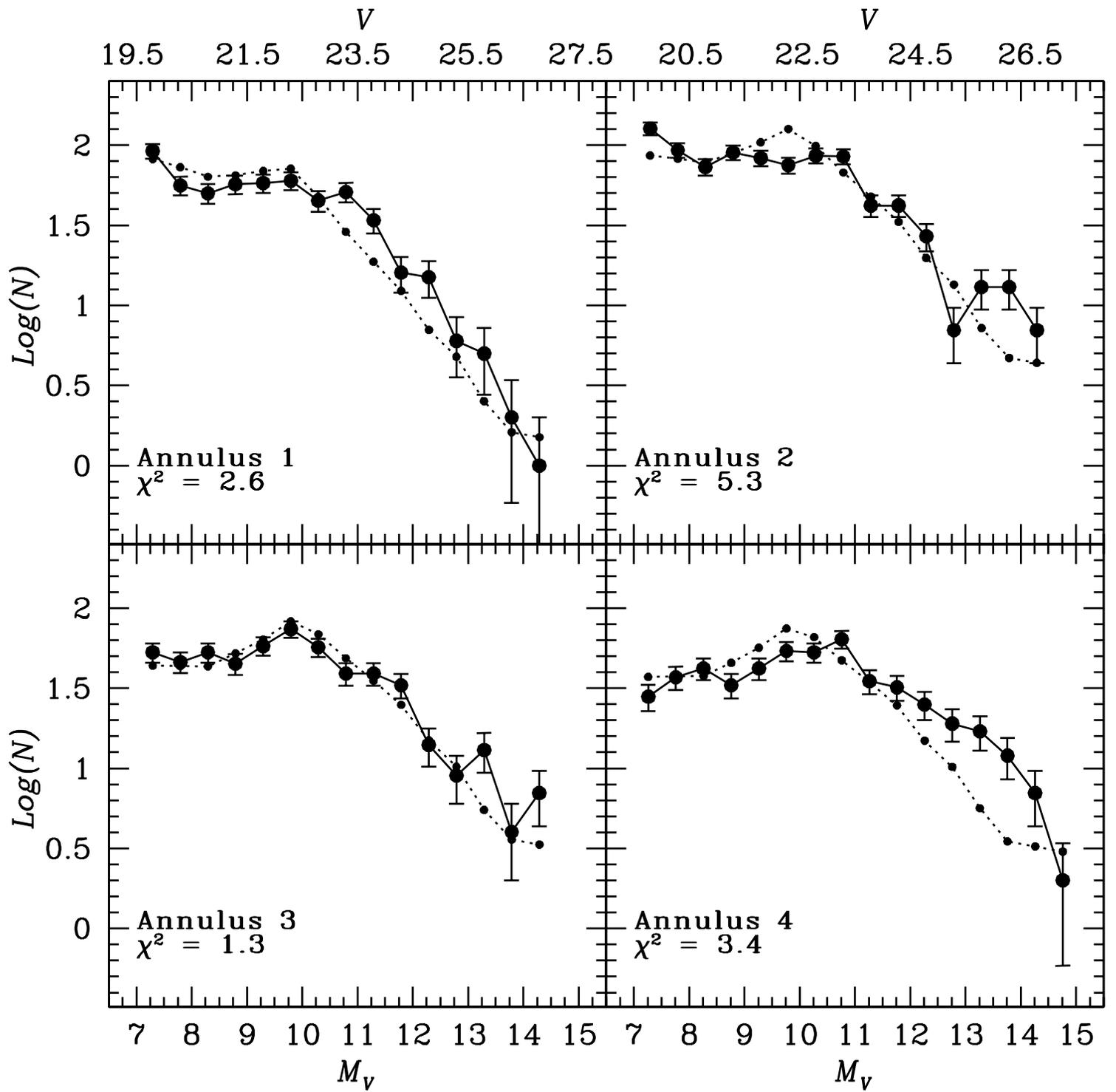
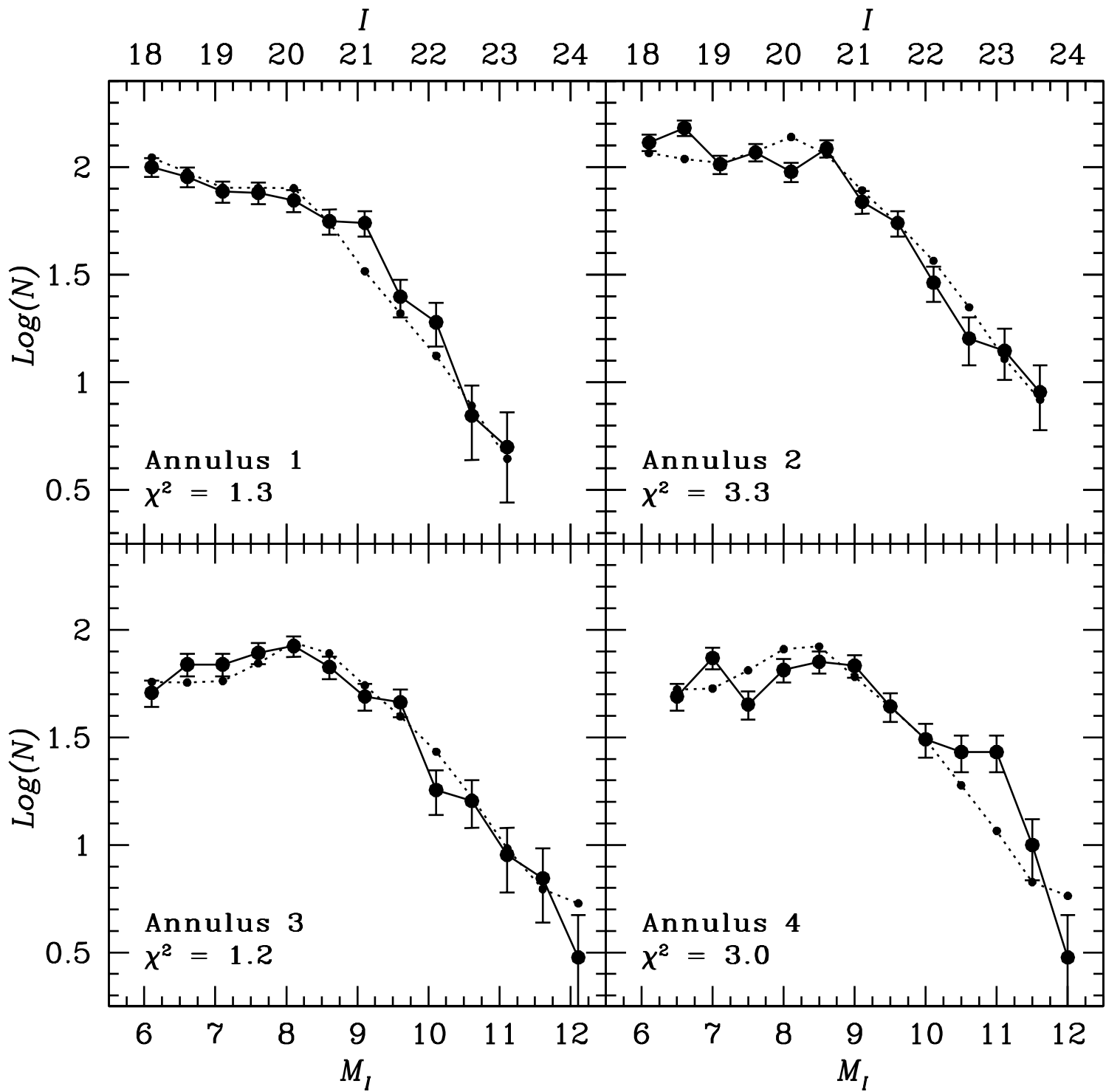
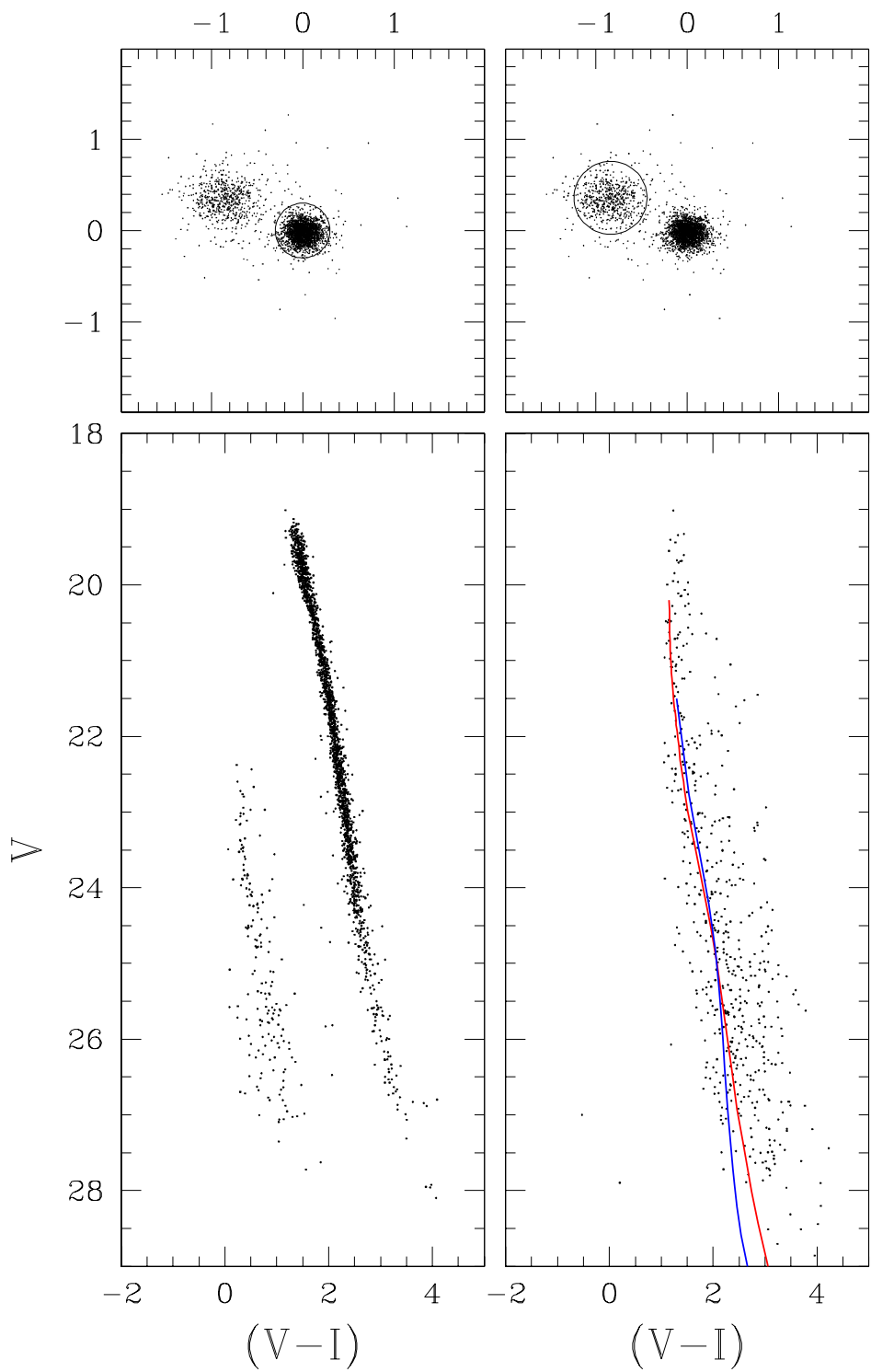
\end{document}